\documentclass[aps,prd,showkeys,showpacs,amssymb,
twocolumn,
amsfonts,epsf,preprintnumbers,nofootinbib,superscriptaddress]{revtex4}

\usepackage[dvips]{graphicx}
\usepackage{bm,latexsym,amsmath,amssymb,amsfonts,color}
\def\be {\begin{equation}}
\def\ee {\end{equation}}
\def\ba {\begin{eqnarray}}
\def\ea {\end{eqnarray}}

\def\bi {\begin{itemize}}
\def\ei {\end{itemize}}
\newcommand\beq{\begin{eqnarray}}
\newcommand\eeq{\end{eqnarray}}
\newcommand{\bea}{\begin{eqnarray}}
\newcommand{\eea}{\end{eqnarray}}

\usepackage{graphicx}% Include figure files
\usepackage{dcolumn}% Align table columns on decimal point
\usepackage{bm,morefloats,color}% bold math

\usepackage{amssymb,bm}

\def\X5sp{{\rm X}_5}
\def\Y3sp{{\rm Y}_3}
\def\Z3sp{{\rm Z}_3}

\begin{document}

\title{Solutions in the scalar-tensor theory with nonminimal derivative coupling}

\author{Masato Minamitsuji}
\affiliation{Multidisciplinary Center for Astrophysics (CENTRA), Instituto Superior T\'ecnico, Lisbon 1049-001, Portugal.}

\begin{abstract}
We present black hole type solutions in the scalar-tensor theory with nonminimal derivative coupling to the Einstein tensor. The effects of the nonminimal derivative coupling appear in the large scales, while the solutions approach those in the Einstein gravity in the small scales. For the particular coupling constant tuned to the inverse of the  cosmological constant,  the scalar field becomes trivial and the solutions in the Einstein gravity are recovered.  For the other coupling constant, more general solutions can be obtained. If the two-dimensional maximally symmetric space is a two-sphere, the spacetime structure approaches anti-de Sitter spacetime in the large scales. On the other hand, if the two-dimensional space is a two-hyperboloid, the spacetime approaches de Sitter (AdS) spacetime. If it is a flat space, the cosmological constant affects only the amplitude of the scalar field. The extension to the higher-dimensional case is also straightforward. For a certain range of the negative cosmological constant, thermodynamic properties of a black hole are very similar to those of the Schwarzschild-AdS black hole in the Einstein gravity.
\end{abstract}

\date{\today}

\maketitle

%%%%%%%%%%%%%%%%%%%%%%%%%
\section{Introduction}

Recently,
various modified gravity theories have been proposed
in the context of presenting successful cosmological models for 
inflation and/or dark energy (see e.g., \cite{mg} and references therein).
Many of these theories can be 
described by the so-called (generalized) Galileon scalar-tensor theories \cite{hor,nic,def,def2,def3,van,kyy,char}.
Among these theories,
we focus on the scalar-tensor theory
with nonminimal derivative coupling to the Einstein tensor
\bea
\label{action}
S&=&\frac{1}{2}
\int d^4x\sqrt{-g}
\Big[
m_p^2
\big(R-2\Lambda\big)
\nonumber\\
&-&
\big(g^{\mu\nu}-
\frac{z}{m_p^2} G^{\mu\nu}\big)
\partial_\mu \phi
\partial_\nu \phi
\Big],
\eea
where the metric $g_{\mu\nu}$ is the metric,
$g={\rm det(g_{\mu\nu})}$,
and $R$ and $G_{\mu\nu}$
are the Ricci scalar and the Einstein tensor for the metric $g_{\mu\nu}$, respectively.
$z$ parametrizes the nonminimal derivative coupling and 
$\Lambda$ is the cosmological constant.
$m_p$ is the reduced Planck mass and
we assume $z= O(1)$.
Despite the higher derivative coupling in the action,
the highest derivative terms in the equations of motion 
are of the second order
because of the contracted Bianchi identity $\nabla_\mu G^{\mu\nu}=0$,
where $\nabla_\mu$ is the covariant derivative with respect to $g_{\mu\nu}$.
%Such a nonminimal derivative coupling may appear
%as one of the leading order corrections to the familiar nonminimal
%coupling to the Ricci scalar $\xi R\phi^2$.
%Since the nonminimal derivative coupling term is suppressed by $m_p^{-2}$,
%the action \eqref{action} is realized
%when the standard nonminimal coupling term vanishes, $\xi=0$,
%and any other higher order derivative coupling term is still subdominant.
In the rest, we will set $m_p=1$ unless it should be shown explicitly.
We will focus on the vacuum solutions in the scalar-tensor theory \eqref{action},
and will not consider solutions with the matter fields except for the scalar field.

The theory given by the action \eqref{action}  is equivalent to the following class of the generalized Galileon scalar-tensor 
theory 
which is constituted of the Lagrangians $L_{i}$ ($i=2,3,4,5$) \cite{def3,kyy}  with
\bea
\label{simplest}
K(\phi,X)&=&X-\Lambda,
\quad
G_3(\phi,X)=0,
\nonumber\\
G_4(\phi,X)&=&\frac{1}{2},
\quad
G_5(\phi,X)=-\frac{z}{2}\phi,
\eea
where we have defined $X:=-\frac{1}{2}g^{\mu\nu}\partial_\mu \phi \partial_\nu\phi$
and followed the definition of (2.1)-(2.4) in \cite{kyy}.
This can be explicitly confirmed by the partial integration of the nonminimal derivative coupling term
in \eqref{action} with the use of the contracted Bianchi identity $\nabla_\mu G^{\mu\nu}=0$,
and by neglecting the total derivative term which does not contribute to the equations of motion.
If we choose $G_5=\rm const$, 
the ${\cal L}_5$ term just reduces to the total derivative term.
Hence, 
$G_5\propto\phi$ in \eqref{simplest} corresponds to one of the simplest choices of the Lagrangian ${\cal L}_5$
which can provide the nontrivial contribution to the equations of motion.
Therefore, the investigation of the theory \eqref{action} 
would be the very important first step to understand the properties of the 
gravitational physics in the scalar-tensor theory with the Lagrangian ${\cal L}_5$.
Taking the nonminimal derivative coupling to the Einstein tensor into consideration 
is also motivated by the low energy effective action of string theory \cite{st,st2,st3}
%%%%%%%%%%%%%%%%%%%%%%%%%%%%%%%%%%%%%%%%%%
%Such coupling also arises from
and the ghost-free nonlinear massive gravity \cite{nlm}.
%%%%%%%%%%%%%%%%%%%%%%%%%%%%%%%%%%%%%%%%%%%%
On the other hand, the cosmological constant $\Lambda$ is equivalently seen as the constant potential term of the scalar field.
The potential term would affect the dynamics of the scalar field and the spacetime metric,
as it modifies the asymptotic structure of the black hole spacetime
and the expansion law of the universe in the Einstein gravity.
%%%%%%%%%%%%%%%%%%%%%%%%%%%%%%%%%%%%%%%%%%%%%%%%%%%%%%%%%
In our present case, it is very interesting to investigate 
how the cosmological constant $\Lambda$ interacts with the nonminimal derivative coupling term at the nonlinear level.
%%%%%%%%%%%%%%%%%%%%%%%%%%%%%%%%%%%%%%%%%%%%%%%%%%%%%%%%%%
In general, in order to understand the properties of the given gravitational theory,
the investigation of exact solutions is particularly important.
From the above points of view,
in this paper we will investigate the exact solutions in the theory \eqref{action}.

This theory has attracted much interest from the cosmological points of view \cite{ame,cap,sch,sar,gra,gra2,cd1,cd2,cd3,ger,ger2,gw,sal,papa,qiu,sarida}.
The accelerating cosmological solutions with the nonminimal derivative coupling
to the curvature  
were first considered in \cite{ame,cap}.
Exact cosmological solutions with derivative coupling to 
the Einstein tensor have been obtained in \cite{sch,sar,gra,gra2}.
Cosmological dynamics in this theory
has been investigated in \cite{cd1,cd2,cd3}.
Slow-roll inflationary models
have been developed in \cite{ger,ger2,gw} as an extension of the nonminimal Higgs inflation model \cite{bez}.
Reheating \cite{sal,papa} and curvaton \cite{qiu} mechanisms
have also been discussed recently.
The linear cosmological perturbation theory in the cosmological scenario with nonminimal derivative coupling
to the Einstein tensor and comparison of it with the observational data have been recently argued in \cite{sarida}.

Black holes would provide another interesting arena to probe the aspects of the modified gravity theories.
References \cite{cj,koly,koly2} investigated
the behavior of a scalar field   derivative coupled to the Einstein tensor
on a charged black hole.
Some exact solutions were obtained via dimensional reduction from the higher-dimensional 
Einstein-Gauss-Bonnet gravity \cite{char}.
A no-hair theorem was argued in \cite{af1}.
A spherically symmetric, static solution 
in the theory \eqref{action} with a vanishing cosmological constant $\Lambda=0$ 
was obtained in \cite{rinaldi}.
%The spacetime structure is very similar to that of the Schwarzschild anti-de Sitter (AdS)  spacetime,
In this solution, the spacetime is asymptotically anti-de Sitter (AdS)  
where the nonminimal derivative coupling plays the role of  a negative cosmological constant.
In this work,
we explore the black hole type solutions
in the case with a nonzero cosmological constant given in \eqref{action}.
The generalization to the case of a nonzero cosmological constant is straightforward.
We also investigate the existence of singularities and the asymptotic behaviors.
Finally, we explore the solutions in the five-dimensional extension of \eqref{action}.
The obtained solutions have very similar properties to their four-dimensional counterparts.
In addition to the spherically symmetric solutions,
our analysis will also involve the solutions with the two-dimensional flat space and those with the two-dimensional hyperbolic space.

The important aim of this paper is to investigate
how adding the cosmological constant 
to the %gravitational theory
scalar-tensor theory with the nonminimal derivative coupling to the Einstein tensor
modifies the properties of the black hole spacetime,
compared to the original solution \cite{rinaldi} obtained without the cosmological constant.
Our investigation 
will reveal nontrivial modifications of the properties of the spacetime.
We will find that even if a positive cosmological constant is added to the theory, 
the black hole spacetime still approaches AdS at the asymptotic infinity.
Moreover, we will also find that adding a positive cosmological constant
larger than the inverse of the coupling constant 
induces a curvature singularity at the finite coordinate position,
and for the particular coupling constant tuned to the inverse of the cosmological constant
the scalar field becomes trivial and hence the black hole solution in the Einstein gravity is recovered. 
These properties will imply the nontrivial interplay of the cosmological constant 
with the nonminimal derivative coupling to the Einstein tensor
in our gravitational theory \eqref{action}.
Thermodynamic properties of a black hole will also be investigated,
which can be very similar to those of the Schwarzschild-AdS spacetime,
depending on the choice of the coupling constant.
In the next sections, we will see these points more explicitly.

This paper is organized as follows:
In Sec. II, we present the equations of motion.
In Sec. III, we discuss solutions with a vanishing cosmological constant.
In Sec. IV, we discuss solutions with a  nonzero cosmological constant
and compare both cases.
In Sec. V, we discuss thermodynamic properties of our black hole solutions
with a cosmological constant. 
In Sec. VI, we present the solutions in the five-dimensional model.
In Sec. VII, we close this paper after giving a brief summary.

%%%%%%%%%%%%%%%%%%%%%%%%%%%%%%%%%%%%%%%%%%%%%%%%%%%%%%%%%%%%%%%%%%%
\section{The equations of motion}

Let us derive the equations of motion by varying the action \eqref{action}.
Varying the action \eqref{action} with respect to the metric $g_{\mu\nu}$,
the gravitational equation is given by 
\bea
\label{eins}
G_{\mu\nu}=T_{\mu\nu}-z L_{\mu\nu}-\Lambda g_{\mu\nu},
\eea
where 
\bea
T_{\mu\nu}:=\nabla_\mu\phi \nabla_\nu \phi
          -\frac{1}{2}g_{\mu\nu}\nabla^{\lambda}\phi\nabla_{\lambda}\phi,
\eea
and 
\bea
L_{\mu\nu}&:=&
- \nabla_{\mu}\nabla_{\nu}\phi \Box\phi
+ \nabla_\lambda \big(\nabla_{\mu} \phi\big)\nabla^\lambda\big(\nabla_{\nu} \phi\big)
\nonumber\\
&+& R_{\mu\alpha \nu\beta}　\nabla^\alpha \phi\nabla^\beta \phi
-\frac{1}{2}R \nabla_{\mu} \phi \nabla_{\nu}\phi 
\nonumber\\
&+&2\nabla^\lambda \phi R_{\lambda(\mu} \nabla_{\nu)}\phi
-\frac{1}{2} G_{\mu\nu}  \nabla_\lambda \phi\nabla^{\lambda}\phi
\nonumber\\
&+&g_{\mu\nu}
\Big(
-R^{\alpha\beta}\nabla_\alpha\phi\nabla_\beta\phi
+\frac{1}{2}\big(\Box\phi\big)^2 
\nonumber\\
&-&\frac{1}{2}\nabla_\alpha\nabla_\beta \phi \nabla^\alpha\nabla^\beta\phi
\Big).
\eea
Similarly, varying the action \eqref{action}
with respect to $\phi$,
the equation of motion of the scalar field is given by
\bea
\label{sca}
\big(g_{\mu\nu}-zG_{\mu\nu}\big)\nabla^\mu\nabla^\nu\phi=0.
\eea
Note that
the highest derivatives are still of the second order
because of $\nabla_\mu G^{\mu\nu}=0$.
In this paper, we will consider the vacuum solutions in the scalar-tensor theory \eqref{action}.
Hence, the energy-momentum tensor 
does not contain the contribution of the matter fields except for the scalar field.

We look for solutions of Eqs. (\ref{eins}) and (\ref{sca})
under the metric ansatz
\bea
\label{ansatz}
ds^2= -f(r) dt^2+ g(r) dr^2+ r^2 d\Omega_K^2,
\eea
where $K=+1,0,-1$ denotes the constant curvature of the 
two-dimensional maximally symmetric space. 
For $f(r)>0$ and $g(r)>0$, $r$ is spacelike
and the spacetime is static,
while for $f(r)<0$ and $g(r)<0$, $r$ is timelike
and the spacetime becomes dynamical.
We also assume that $\phi=\phi(r)$.
The nontrivial components of the gravitational equation \eqref{eins}
are given by 
\bea
\label{eins2}
&&\frac{r}{2}\big(3z(\phi')^2+ 2g\big)
\Big(\frac{f'}{f}-\frac{g'}{g}\Big)
\nonumber\\
&=&2g(Kg-1) -2z (\phi')^2-zr\{ (\phi')^2\}'-2\Lambda r^2 g^2,
\nonumber\\
&&r\big(3z (\phi')^2+2g\big)
\frac{f'}{f}
\nonumber\\
&=&2g (Kg-1)+z(\phi')^2 (Kg-3)+r^2 g(\phi')^2-2\Lambda r^2 g^2,
\nonumber\\
\eea
where a ``prime'' denotes the derivative with respect to $r$.
The scalar field equation of motion \eqref{sca}
is given by 
\bea
\label{kir}
\frac{d}{dr}
\Big[
\frac{f^{\frac{1}{2}}}{g^{\frac{3}{2}}}
\Big(
\frac{rf'}{f}
-\big(Kg-1\big)
-\frac{r^2 g}{z}
\Big)
\phi'
\Big]
=0.
\eea
We integrate Eq. (\ref{kir}) with respect to $r$ and
set the integration constant to be zero along the same line of \cite{rinaldi}.
Then the equation \eqref{kir} can further reduce to
\bea
\label{sca2}
\frac{rf'}{f}=Kg-1+\frac{r^2 g}{z}.
\eea
If we choose a nonzero integration constant,
$f$ is not decoupled from the other variables
and hence to find analytic solutions becomes a more complicated problem. 
Note that  because in the theory \eqref{action} there is the shift symmetry 
under the transformation $\phi\to \phi+{\rm const}$,
we will present the derivative of the scalar field $\phi'$ which can be uniquely determined, instead of $\phi$ itself.

\section{Solutions with a vanishing cosmological constant}

We then solve  Eqs. (\ref{eins2}) and (\ref{sca2}).
In this section, 
we look for the solutions 
of the vanishing cosmological constant $\Lambda=0$.
In the next section, we look for solutions for $\Lambda\neq 0$.

Note that the solutions for $K=+1$ have been obtained in \cite{rinaldi},
while those for $K=-1$ and $K=0$ can also be obtained in the similar way.
An explicit comparison of them with the solutions with a cosmological constant 
will be done in the next section.

\subsection{The solutions for $K=+1$}

The solutions for $K=+1$ are given as follows.

\subsubsection{$z>0$}

For $z>0$, the general solution is given by 
\bea
\label{rina}
f(r)&=&\frac{3}{4}-\frac{2m}{r} +\frac{r^2}{12z}+\frac{\sqrt{z}}{4r}{\rm arctan}
\Big(\frac{r}{\sqrt{z}}\Big),
\nonumber\\
g(r)&=&\frac{(r^2+2z)^2}{4(r^2+z)^2f(r)},
\nonumber\\
(\phi'(r))^2&=&-\frac{(r^3+2rz)^2}{4(r^2+z)^3z f(r)}.
\eea
This solution was obtained in \cite{rinaldi}.
The overall normalization of $f(r)$
is chosen to recover the Schwarzschild black hole solution
in the small $r$ limit.
We obtain
\bea
\label{srp}
f(r)= -\frac{2m}{r}+1+\frac{r^4}{20z^2}+O(r^6),
%+O(r^4)
\eea
and $g(r)\approx 1/f(r)$.
The explicit dependence on the coupling constant $z$ appears in the $O(r^4)$ term,
by which we could distinguish the present model from the Einstein gravity without a cosmological constant. 
The Schwarzschild solution can also be obtained in the limit of $z\to \infty$.
On the other hand,
in the large $r$ limit,
the effects of the nonminimal
derivative coupling become more important and 
we obtain
\bea
f(r)&=&
\frac{r^2}{12 z}+\frac{3}{4}
+\frac{-16m+\pi\sqrt{z}}{8r}
+O(r^{-2}).
\eea
Thus the asymptotic structure in the large $r$ limit is 
AdS spacetime
with the effective cosmological constant $-\frac{1}{4z}(<0)$
\cite{rinaldi}.
The point $f(r)=0$ corresponds to the horizon.
The derivative of the scalar field
with respect to the proper length, $\frac{1}{\sqrt{|g|}}|\phi'(r)|$,
remains finite at the horizon.
Note that $(\phi'(r))^2$ is negative and 
the scalar field becomes ghostlike outside the horizon
because $f(r)>0$ for $r>r_h$,
where $r_h$ is the position of the horizon so that $f(r_h)=0$.
In the next section, we will see 
that this property can be improved by adding a negative cosmological constant.

Following \cite{entropy,rinaldi},
the temperature of the black hole, $\beta^{-1}$, is related to the surface gravity  
$\kappa:=\frac{1}{2}\frac{(-g_{tt})'}{\sqrt{-g_{rr}g_{tt}}}\big|_{r=r_h}$
as $\beta=\frac{2\pi}{\kappa}$,
and
hence
\bea
\beta=\frac{8\pi z r_h }{r_h^2+ 2z}.
\eea
For a large $z$, it reproduces the temperature of the Schwarzschild black hole
$\beta=4\pi r_h$.

%The other important observation of our solution
%is that because $f(r)>0$ for $r>r_h$
%for $\Lambda z<-1$, $(\phi')^2>0$ outside the horizon,
%while for $\Lambda z>-1$ including the case of $z=0$, $(\phi')^2<0$ outside the horizon.
%Thus adding the negative cosmological constant ensures the positively of $(\phi')^2$ outside the horizon.

\subsubsection{$z<0$}

For $z<0$, the general solution is given by 
\bea
f(r)&=&\frac{3}{4}-\frac{2m}{r} +\frac{r^2}{12z}+\frac{\sqrt{-z}}{4r}{\rm arctanh}
\Big(\frac{r}{\sqrt{-z}}\Big),
\nonumber\\
g(r)&=&\frac{(r^2+2z)^2}{4(r^2+z)^2f(r)},
\nonumber\\
(\phi'(r))^2&=&-\frac{(r^3+2rz)^2}{4(r^2+z)^3z f(r)},
\eea
where the domain of the coordinate $r$ is given by $0<r<\sqrt{-z}$.
The overall normalization of $f(r)$
is  chosen to recover the Schwarzschild solution 
in the small $r$ limit
\eqref{srp}.
The Schwarzchild solution can also be obtained in the limit of $z\to -\infty$.
At the boundary $r= \sqrt{-z}$, the spacetime is regular
but the proper distance to the boundary is infinite.
%$(\phi')^2$ is negative outside the horizon $r>r_h$ and positive inside it.

\subsection{The solutions for $K=-1$}

The solutions for $K=-1$ are given as follows.

\subsubsection{$z<0$}

For $z<0$, the general solution is given by 
\bea
\label{rina2}
f(r)&=&-\frac{3}{4}-\frac{2m}{r} +\frac{r^2}{12z}-\frac{\sqrt{-z}}{4r}{\rm arctan}
\Big(\frac{r}{\sqrt{-z}}\Big),
\nonumber\\
g(r)&=&\frac{(r^2-2z)^2}{4(r^2-z)^2f(r)},
\nonumber\\
(\phi'(r))^2&=&-\frac{(r^3-2rz)^2}{4(r^2-z)^3z f(r)}.
\eea
The overall normalization of $f(r)$
is chosen to recover the solution of $K=-1$ in the Einstein gravity
in the small $r$ limit.
We obtain
\bea
\label{srm}
f(r)= -\frac{2m}{r}-1-\frac{r^4}{20z^2}+O(r^6),
%+ O(r^4),
\eea
and $g(r)\approx 1/f(r)$.
The explicit dependence on the coupling constant $z$ appears in the $O(r^4)$ term,
by which we could distinguish the present model from the Einstein gravity without a cosmological constant. 
The solution in the Einstein gravity can also be obtained in the limit of $z\to -\infty$.
In the large $r$ limit,
the effects of the nonminimal derivative coupling become more important and 
we obtain
\bea
f(r)&=&
-\frac{r^2}{12 (-z)}
-\frac{3}{4}
\nonumber\\
&-&\frac{16m+\pi\sqrt{-z}}{8r}
+O(r^{-2}).
\eea
Thus 
the asymptotic structure in the large $r$ limit is the de Sitter (dS) spacetime
with the effective cosmological constant $\frac{1}{4(-z)}(>0)$.
The derivative of the scalar field with respect to the proper length, $\frac{1}{\sqrt{|g|}}|\phi'(r)|$,
remains finite at the horizon.
Note that $(\phi'(r))^2$ is negative and 
the scalar field  becomes ghostlike
in the large-$r$ region because of $f(r)<0$.
In the next section, we will see 
that this property can be improved by adding a positive cosmological constant.

%Since $f(r)$ is negative for $m>0$, $(\phi')^2$ is negative.
%As we will see in the next section,
%adding a cosmological constant can make $(\phi')^2$ positive.

\subsubsection{$z>0$}

For $z>0$, the general solution is given by 
\bea
f(r)&=&-\frac{3}{4}-\frac{2m}{r} +\frac{r^2}{12z}-\frac{\sqrt{z}}{4r}{\rm arctanh}
\Big(\frac{r}{\sqrt{z}}\Big),
\nonumber\\
g(r)&=&\frac{(r^2-2z)^2}{4(r^2-z)^2f(r)},
\nonumber\\
(\phi'(r))^2&=&-\frac{(r^3-2rz)^2}{4(r^2-z)^3z f(r)},
\eea
where the domain of the coordinate $r$ is given by $0<r<\sqrt{z}$.
The overall normalization of $f(r)$
is chosen to recover 
the solution of $K=-1$ in the Einstein gravity
in the small $r$ limit
\eqref{srm}.
The solution in the Einstein gravity can also be obtained in the limit of $z\to \infty$.
At the boundary $r= \sqrt{z}$, the spacetime is regular.
%Since $f(r)$ is negative for $m>0$,  $(\phi')^2$ is negative.

\subsection{The solutions for $K=0$}

The solution for $K=0$ is given by
\bea
\label{hind}
f(r)&=&-\frac{2m}{r} +\frac{r^2}{12z},
\quad
g(r)=\frac{1}{4f(r)},
\nonumber\\
(\phi'(r))^2&=&-\frac{1}{4z f(r)}.
\eea
This solution is singular only at $r=0$.
%In the case of $m>0$, 
%$(\phi')^2$ is always negative for $z<0$
%and it is negative outside the horizon for $z>0$.

%\subsection{Summary}

%The properties of solutions with a vanishing cosmological constant
%are summarized in the TABLE I which will be shown in the subsection IV-D.
%In the table, 
%the domain of the spacetime,
%position of the singularities
%and asymptotic behavior of the spacetime
%are %summarized. 
%listed.
%On the other hand,
%the asymptotic behavior completely depends on the
%sign of $K$ and $z$.
%%%%%%%%%%%%%%%%%%%%%%%%%%%%%%%%%%%%%%%%%%%%%%%%%%%%
%\begin{widetext}
%\begin{center}
%\label{table1}
%\begin{tabular}{|c|c|c|c|c|c|c|c|c|
%}
%\hline
%&\multicolumn{2}{c|}{$K=+1$}
%&\multicolumn{2}{c|}{$K=-1$}
%&\multicolumn{2}{c|}{$K=0$}
%\\ \hline  
% & $z>0$ 
% & $z<0$
% & $z>0$ 
% & $z<0$ 
% & $z>0$ 
% & $z<0$
%\\ \hline\hline
%Domain
%& $0<r<\infty$
%& $0<r<\sqrt{-z}$
%& $0<r<\sqrt{z}$
%& $0<r<\infty$
%& $0<r<\infty$
%& $0<r<\infty$ 
 % \\ \hline
 %Singularites
%& $r=0$
%& $r=0$
%& $r=0$
%& $r=0$
%& $r=0$
%& $r=0$
% \\ \hline
%Asymptotic Behavior
%& AdS 
%& -
%& -
%& dS
%& AdS 
%& dS
% \\ \hline
%\end{tabular}
%\end{center}
%\end{widetext}
%%%%%%%%%%%%%%%%%%%%%%%%%%%%%%%%%%%%%%%%%%%%%%%%%%%%%%

\section{Solutions with a cosmological constant}

We then present the solutions with a nonzero cosmological constant $\Lambda\neq 0$.

\subsection{The solutions for $K=+1$}

The solutions for $K=+1$ are given as follows.

\subsubsection{$z>0$ ($z\neq -\frac1\Lambda$)}

For $z>0$, the general solution is given by 
\bea
\label{solp}
f(r)&=&\frac{1}{12rz}
\Big(-24m z+r^3(1-\Lambda z)^2
\nonumber\\
&-&3r z(-1+\Lambda z)(3+\Lambda z)
\nonumber\\
&+&
3z^{3/2} (1+\Lambda z)^2{\rm arctan}\big(\frac{r}{\sqrt{z}}\big)
\Big),
\nonumber\\
g(r)&=&\frac{\big(-2z +r^2(-1+\Lambda z)\big)^2}
                {4(r^2+z)^2f(r)},
\nonumber\\
(\phi'(r))^2&=&-\frac{(1+\Lambda z)\big(-2r z+r^3 (-1+\Lambda z)\big)^2}
                 {4(r^2+z)^3z f(r)},
\eea
where the domain of the coordinate $r$ is given by $0<r<\infty$.
The overall normalization of $f(r)$
is chosen to recover 
the Schwarzschild (A)dS
solution in the small $r$ limit.
We obtain
\bea
\label{srpc2}
f(r)= -\frac{2m}{r}+1-\frac{\Lambda r^2}{3}%+O(r^4),
+\frac{(1+\Lambda z)^2r^4}{20z^2}
+O(r^6)
\eea
and $g(r)\approx 1/f(r)$.
The explicit dependence on the coupling constant $z$ appears in the $O(r^4)$ term,
by which we could distinguish the present model from the Einstein gravity with a cosmological constant. 
Note that the $z$-dependent correction vanishes in the limit $z\to -\frac{1}{\Lambda}$
which corresponds to the case discussed in Sec. IV A 3. 
In the large $z$ limit, we obtain
\bea
\label{largez}
f(r)&=& 1-\frac{2m}{r}-\frac{\Lambda }{3}r^2 +\frac{\Lambda^2}{20}r^4
+O(\frac{1}{z}),
\nonumber\\
f(r)g(r)&=&\Big(1-\frac{r^2\Lambda}{2}\Big)^2+O(\frac{1}{z}).
\eea
In the large $r$ limit,
the effects of the nonminimal derivative coupling become more important and 
we obtain
\bea
\label{lr1}
f(r)&=&
\frac{r^2 (\Lambda  z-1)^2}{12 z}
-\frac{1}{4} (\Lambda  z-1) (\Lambda  z+3)
\nonumber\\
&+&\frac{-16m+\pi\sqrt{z} (1+\Lambda z)^2}{8r}
+O(r^{-2}).
\eea
Thus the asymptotic structure in the large $r$ limit is 
AdS spacetime
with the effective cosmological constant $-\frac{(1-\Lambda z)^2}{4z} (<0)$.
The horizon is formed at the place where $f(r)=0$.
The event horizon is always formed
but no cosmological horizon is formed
even for $\Lambda>0$,
because of the asymptotically AdS property.
Another interesting property obtained by adding a cosmological constant $\Lambda$
is that 
$(\phi'(r))^2$ can be positive outside the horizon for $\Lambda <-\frac{1}{z}$
and the scalar field does not become ghostlike,
because $f(r)>0$ for $r>r_h$,
where $r_h$ is the position of the horizon so that $f(r_h)=0$.

The derivative of the scalar field with respect to the proper length, $\frac{1}{\sqrt{|g|}}|\phi'(r)|$,
remains finite at the horizon.
As discussed in the previous section,
the temperature of the black hole  $\beta^{-1}$ is given by 
\bea
\label{temp_lam}
\beta=\frac{8\pi z r_h }{r_h^2\big(1-\Lambda z\big)+ 2z}.
\eea
For a large $z$ and a fixed finite $\Lambda z$,
 it reproduces the temperature of the Schwarzschild black hole
$\beta=4\pi r_h$.
On the other hand, for both large $z$ and $-\Lambda z$ ($\Lambda<0$),
$\beta=\frac{4\pi  r_h }{1-\frac{\Lambda}{2}r_h^2}$,
which does not agree with the temperature of the Schwarzschild AdS black hole
$\beta=\frac{4\pi  r_h }{1-\Lambda r_h^2}$,
because \eqref{largez} is not precisely the same
as the metric of the Schwarzschild AdS.
The point of $g(r)=0$,
$r=\sqrt{\frac{2z}{\Lambda z-1}}$,
becomes a singularity
because the invariant  $R^{\alpha\beta\mu\nu}R_{\alpha\beta\mu\nu}$ diverges there.
This point appears at a finite coordinate position for $\Lambda>\frac{1}{z}$,
while for $\Lambda<\frac{1}{z}$ there is no such singularity.
%The singularity at $r=\sqrt{\frac{2z}{\Lambda z-1}}$ is hidden by the horizon.
%On the other hand,
%for $\Lambda<-\frac{1}{z}$,
%$(\phi')^2$ becomes positive outside the horizon $r>r_h$.

\subsubsection{$z<0$ ($z\neq -\frac1\Lambda$)}

For $z<0$, the general solution is given by 
\bea
f(r)&=&\frac{1}{12rz}
\Big(-24m z+r^3(1-\Lambda z)^2
\nonumber\\
&-&3r z(-1+\Lambda z)(3+\Lambda z)
\nonumber\\
&+&3z (1+\Lambda z)^2
(-z)^{\frac{1}{2}} {\rm arctanh}\big(\frac{r}{\sqrt{-z}}\big)
\Big),
\nonumber\\
g(r)&=&\frac{\big(-2z +r^2(-1+\Lambda z)\big)^2}
                {4(r^2+z)^2f(r)},
\nonumber\\
(\phi'(r))^2&=&-\frac{(1+\Lambda z)\big(-2r z+r^3 (-1+\Lambda z)\big)^2}
                 {4(r^2+z)^3z f(r)},
\eea
where the domain of the coordinate $r$ is given by $0<r<\sqrt{-z}$.
The overall normalization of $f(r)$
is chosen to recover 
the Schwarzschild-(A)dS solution 
in the small $r$ limit \eqref{srpc2}.
In the large $(-z)$ limit, we reproduce \eqref{largez}.
At the boundary $r= \sqrt{-z}$, the spacetime is regular
but the proper distance to the boundary is infinite.
For $\Lambda>\frac{1}{(-z)}$,
a singularity appears at $r=\sqrt{\frac{2(-z)}{1+\Lambda (-z)}} (<\sqrt{-z})$.
%For $\Lambda>\frac{1}{(-z)}$,
%$(\phi')^2$ becomes positive outside the horizon $r>r_h$.
%The singularity at $r=\sqrt{\frac{2(-z)}{1+\Lambda (-z)}}$ is not hidden by the horizon.

\subsubsection{$z=-\frac{1}{\Lambda}$}

For $z=-\frac{1}{\Lambda}$, 
the solution is given by
\bea
f(r)&=&1-\frac{2m}{r}-\frac{\Lambda}{3}r^2,\quad
g(r)=\frac{1}{f(r)},
\nonumber\\
\phi'(r)&=&0.
\eea
The scalar field  becomes trivial and the 
Schwarzschild-(A)dS solution is recovered.

\subsection{The solutions for $K=-1$}

The solutions for $K=-1$ are given as follows.

\subsubsection{$z<0$ ($z\neq -\frac1\Lambda$)}

For $z<0$, the general solution is given by 
\bea
\label{soln}
f(r)&=&\frac{1}{12rz}
\Big(-24m z+r^3(1-\Lambda z)^2
\nonumber\\
&+&3r z(-1+\Lambda z)(3+\Lambda z)
\nonumber\\
&-&3z (1+\Lambda z)^2
(-z)^{1/2} {\rm arctan}\big(\frac{r}{\sqrt{-z}}\big)
\Big),
\nonumber\\
g(r)&=&\frac{\big(2z +r^2(-1+\Lambda z)\big)^2}
                {4(r^2-z)^2f(r)},
\nonumber\\
(\phi'(r))^2&=&-\frac{(1+\Lambda z)\big(2r z+r^3 (-1+\Lambda z)\big)^2}
                 {4(r^2-z)^3z f(r)},
\eea
where the domain of the coordinate $r$ is given by $0<r<\infty$.
The overall normalization of $f(r)$
is chosen to recover
the solution of $K=-1$ in the Einstein gravity
in the small $r$ limit.
We obtain
\bea
\label{srmc}
f(r)= -\frac{2m}{r}-1-\frac{\Lambda r^2}{3}-\frac{(1+\Lambda z)^2r^4}{20z^2}+O(r^6),
%+O(r^4),
\eea
and $g(r)\approx 1/f(r)$.
The explicit dependence on the coupling constant $z$ appears in the $O(r^4)$ term,
by which we could distinguish the present model from the Einstein gravity with a cosmological constant. 
Note that the $z$-dependent correction vanishes in the limit $z\to -\frac{1}{\Lambda}$
which corresponds to the case discussed in Sec. IV B 3. 
In the large $(-z)$ limit, we obtain
\bea
\label{largez2}
f(r)&=& -1-\frac{2m}{r}-\frac{\Lambda }{3}r^2 -\frac{\Lambda^2}{20}r^4
+O(\frac{1}{(-z)}),
\nonumber\\
f(r)g(r)&=&\Big(1+\frac{r^2\Lambda}{2}\Big)^2+O(\frac{1}{(-z)}).
\eea
In the large $r$ limit,
the effects of the nonminimal derivative coupling become more important and 
we obtain
\bea
\label{lr2}
f(r)&=&
\frac{r^2 (\Lambda  z-1)^2}{12 z}
+\frac{1}{4} (\Lambda  z-1) (\Lambda  z+3)
\nonumber\\
&-&\frac{16m+\pi\sqrt{-z} (1+\Lambda z)^2}{8r}
+O(r^{-2}).
\eea
Thus the asymptotic structure in the large $r$ limit is dS spacetime
with the effective cosmological constant $\frac{(1-\Lambda z)^2}{4(-z)} (>0)$.
For $m<0$, $f(r)$ is always negative,
and $t$ and $r$ are always spacelike and timelike, respectively.

The derivative of the scalar field with respect to the proper length, $\frac{1}{\sqrt{|g|}}|\phi'(r)|$,
remains finite at the horizon.
On the other hand, the point $g(r)=0$, $r=\sqrt{\frac{2(-z)}{(-\Lambda)(-z)-1}}$,
becomes singularity
because the invariant  $R^{\alpha\beta\mu\nu}R_{\alpha\beta\mu\nu}$ diverges there.
This point appears at a finite coordinate position for $\Lambda<-\frac{1}{(-z)}$
while for $\Lambda>-\frac{1}{(-z)}$
there is no such singularity.
Another interesting property obtained by adding a cosmological constant $\Lambda$
is that 
$(\phi'(r))^2$ can be positive 
in the large-$r$ region
and the scalar field does not become ghostlike 
if $\Lambda>\frac{1}{(-z)}$ because of $f(r)<0$.
%because $f(r)>0$ for $r>r_h$,
%where $r_h$ is the position of the horizon so that $f(r_h)=0$.}

\subsubsection{$z>0$ ($z\neq -\frac1\Lambda$)}

For $z>0$, the general solution is given by 
\bea
f(r)&=&\frac{1}{12rz}
\Big(-24m z+r^3(1-\Lambda z)^2
\nonumber\\
&+&3r z(-1+\Lambda z)(3+\Lambda z)
\nonumber\\
&-&3z^{3/2}(1+\Lambda z)^2
 {\rm arctanh}\big(\frac{r}{\sqrt{z}}\big)
\Big),
\nonumber\\
g(r)&=&\frac{\big(2z +r^2(-1+\Lambda z)\big)^2}
                {4(r^2-z)^2f(r)},
\nonumber\\
(\phi'(r))^2&=&-\frac{(1+\Lambda z)\big(2r z+r^3 (-1+\Lambda z)\big)^2}
                 {4(r^2-z)^3z f(r)},
\eea
where the domain of the coordinate $r$ is given by $0<r<\sqrt{z}$.
The overall normalization of $f(r)$
is chosen to recover 
 the solution in the Einstein gravity
in the small $r$ limit \eqref{srmc}.
In the large $z$ limit, we reproduce \eqref{largez2}.
At the boundary $r= \sqrt{z}$, the spacetime is regular.
For $\Lambda<-\frac{1}{z}$, 
a singularity appears at $r=\sqrt{\frac{2z}{1-\Lambda z} }(<\sqrt{z})$.
%For $\Lambda<-\frac{1}{z}$ and $m>0$, $(\phi')^2$ becomes always positive.

\subsubsection{$z=-\frac{1}{\Lambda}$}

For $z=-\frac{1}{\Lambda}$,  
the solution is given by
\bea
f(r)&=&-1-\frac{2m}{r}-\frac{\Lambda}{3}r^2,\quad
g(r)=\frac{1}{f(r)},
\nonumber\\
\phi'(r)&=&0.
\eea
Thus the scalar field becomes trivial.

\subsection{The solutions for $K=0$}

\subsubsection{$z\neq -\frac{1}{\Lambda}$}

The solution for $K=0$ with $z\neq-\frac{1}{\Lambda}$ is given by
\bea
\label{mond}
f(r)&=&-\frac{2m}{r} +\frac{r^2}{12z},
\quad
g(r)=\frac{1}{4f(r)},
\nonumber\\
(\phi'(r))^2&=&-\frac{1+\Lambda z}{4z f(r)}.
\eea
Thus the dependence on the cosmological constant does not
explicitly appear in the metric functions \cite{sar}.
This solution is singular only at $r=0$.
%For $\Lambda<-\frac{1}{z}$ and $m>0$, $(\phi')^2$ becomes positive
%while for $\Lambda>-\frac{1}{z}$ it is negative.

For $m=0$
and $z<0$,
the solution \eqref{mond} can be rewritten into the form of the
flat Friedmann-Lema\^itre-Robertson-Walker  metric
\bea
ds^2&=& -d\tau^2 +e^{2H_z\tau }d{\bf x}^2,
\quad 
\phi(\tau)=\sqrt{\frac{1+\Lambda z}{z}}\tau.
\eea
where $H_z:=\sqrt{-\frac{1}{3z}}$ 
and $\tau$ is the proper time coordinate 
defined as $r=e^{H_z\tau}$, 
which agrees with the corresponding solution obtained in \cite{sar}.
%The scalar field becomes real for $\Lambda>\frac{1}{(-z)}$.

For $m<0$ and $z<0$, 
by introducing the new coordinates $r=\frac{2|m|^{\frac{1}{3}}}{H_z}\cosh^{\frac{2}{3}} \Big(\frac{3H_z\tau}{2}\Big)$
and $t=\frac{2}{H_z} (x^3)$
into the form of the Bianchi-I  metric
\bea
\label{bianchi}
ds^2&=&-d\tau^2
+\frac{2^{\frac{4}{3}} |m|^{\frac{2}{3}}}
         {H_z^{\frac{4}{3}}}
\sinh^{\frac{2}{3}}(3H_z \tau)
\nonumber\\
&\times&
\Big[
\tanh^{\frac{4}{3}}
\Big(\frac{3}{2}H_z \tau\Big)(dx^3)^2
+
\frac{d\Omega_{K=0}^2}
       {\tanh^{\frac{2}{3}}
\Big(\frac{3}{2}H_z \tau\Big)}
\Big].
\eea
The initial behavior around $\tau\sim 0$ is the same as  the regular branch of 
the Kasner solution \cite{kasner},
and then the universe approaches dS spacetime
in the late time limit \cite{gcp,km}.

\subsubsection{$z=-\frac{1}{\Lambda}$}

For $z=-\frac{1}{\Lambda}$,
the derivative of the scalar field becomes trivial.
In case of $m=0$,
depending on the sign of $\Lambda$,
the solution becomes either dS or AdS.

\subsection{Comparison with the case without a cosmological constant}

In this subsection, we compare the solutions with a cosmological constant
with those without it.
In Tables I and II,
we have listed the basic properties of the solutions.
% without and with a cosmological constant, respectively.
``Domain'', ``Singularities'' and ``Asymptotic behavior'' in these tables correspond to 
the domain of the $r$ coordinate, the position of the curvature singularities,
and the asymptotic behavior of spacetime in the $r\to \infty$ limit,  respectively.
The modifications due to a finite cosmological constant 
appear in the various properties of the spacetime. 
%the solutionshave not been expected a priori.
%These may be thought due to the nontrivial interactions of the nonminimal derivative coupling and 
%cosmological constant in the action \eqref{action}.

%%%%%%%%%%%%%%%%%%%%%%%%%%%%%%%%%%%%%%%%%%%%%%%%%%%%
\begin{widetext}
\begin{table}[h]
\begin{minipage}[t]{.80\textwidth}
\label{table1}
\caption{
The properties of solutions without a cosmological constant.}
\begin{center}
\begin{tabular}{|c|c|c|c|c|c|c|c|c|
}
\hline
&\multicolumn{2}{c|}{$K=+1$}
&\multicolumn{2}{c|}{$K=-1$}
&\multicolumn{2}{c|}{$K=0$}
\\ \hline  
 & $z>0$ 
 & $z<0$
 & $z>0$ 
 & $z<0$ 
 & $z>0$ 
 & $z<0$
\\ \hline\hline
Domain
& $0<r<\infty$
& $0<r<\sqrt{-z}$
& $0<r<\sqrt{z}$
& $0<r<\infty$
& $0<r<\infty$
& $0<r<\infty$ 
  \\ \hline
 Singularities
& $r=0$
& $r=0$
& $r=0$
& $r=0$
& $r=0$
& $r=0$
 \\ \hline
Asymptotic behavior
& AdS 
& -
& -
& dS
& AdS 
& dS
 \\ \hline
\end{tabular}
\end{center}
\end{minipage}
%%%%%%%%%%%%%%%
%\begin{table}[h]
\begin{minipage}[t]{.80\textwidth}
\caption{
The properties of solutions with a cosmological constant
}
\begin{center}
\label{table2}
\begin{tabular}{|c|c|c|c|c|c|c|c|c|
}
\hline
&\multicolumn{2}{c|}{$K=+1$}
&\multicolumn{2}{c|}{$K=-1$}
&\multicolumn{2}{c|}{$K=0$}
\\ \hline  
 & $z>0$ 
 & $z<0$
 & $z>0$ 
 & $z<0$ 
 & $z>0$ 
 & $z<0$
\\ \hline\hline
Domain
& $0<r<\infty$
& $0<r<\sqrt{-z}$
& $0<r<\sqrt{z}$
& $0<r<\infty$
& $0<r<\infty$
& $0<r<\infty$ 
  \\ \hline
 Singularities
& $r=0$
& $r=0$
& $r=0$
& $r=0$
& $r=0$
& $r=0$
 \\ %\hline
% S
& $r=\sqrt{\frac{2z}{\Lambda z-1}}$
& $r=\sqrt{\frac{2(-z)}{1+\Lambda (-z)}}$
& $r=\sqrt{\frac{2z}{1-\Lambda z}}$
& $r=\sqrt{\frac{2(-z)}{\Lambda z-1}}$
& 
& 
\\ 
&  ($\Lambda>\frac{1}{z}$)
&   ($\Lambda>\frac{1}{(-z)}$)
&  ($\Lambda<-\frac{1}{z}$)
&  ($\Lambda<-\frac{1}{(-z)}$)
& 
&
\\
\hline
Asymptotic behavior
& AdS 
& -
& -
& dS
& AdS 
& dS
\\ \hline
\end{tabular}
\end{center}
\end{minipage}
\end{table}
\end{widetext}
%%%%%%%%%%%%%%%%%%%%%%%%%%%%%%%%%%%%%%%%%%%%%%%%%%%%%

One of the most important properties which is absent 
in the case without a cosmological constant
is that the black hole solutions in the Einstein gravity are exactly recovered 
if we choose the particular coupling constant  $z=-\frac{1}{\Lambda}$.
This is the case that the scalar field becomes trivial $\phi'(r)=0$ at any position $r$,
as it is clear from Eqs. \eqref{solp}, \eqref{soln}, and \eqref{mond}.
This is surprising,
because the solutions in the Einstein gravity 
are recovered for the finite coupling constant $z=-\frac{1}{\Lambda}$, 
not for the vanishing coupling constant $z=0$ as we naively guess.
In addition to the fact that 
if $\Lambda=0$,
from Eqs. \eqref{rina}, \eqref{rina2}, and \eqref{hind},
the scalar field becomes trivial and the solutions in the Einstein gravity are recovered 
in the limit of $|z|\to \infty$, 
this also gives another indication that $z$ should  be a nonperturbative parameter
along the line argued in \cite{rinaldi}.

For the black hole case $K=+1$ and $z>0$, adding a negative cosmological constant can also prevent the scalar field
from being ghostlike outside the horizon.
We also find that adding a negative cosmological constant can make
the effective energy-momentum tensor of the scalar field
including the contribution of the nonminimal derivative coupling
obtained from the right-hand side of \eqref{eins}
(but without the contribution of the cosmological constant)
\bea
\label{abenida}
\tilde T^{\mu}{}_{\nu}:=T^{\mu}{}_{\nu}-z L^{\mu}{}_{\nu}
=
{\rm diag} (-\tilde\rho,\tilde p_r,\tilde p_a),
\eea
where 
$\tilde \rho$ is the effective energy density,
and $\tilde p_r$ and $\tilde p_a$ 
are the effective pressures in the radial direction and in the direction of the two-sphere,
respectively,
to satisfy the weak energy condition for which any timelike observer measures a local positive energy density:  
\bea
\label{kyoko}
\tilde\rho\geq 0, 
\quad
\tilde\rho+\tilde p_r\geq 0,
\quad
\tilde\rho+\tilde p_a\geq 0,
\eea
outside the horizon.
In the case of $\Lambda<\frac{1}{z}$, the solution is regular except for $r=0$.
From \eqref{solp} [and \eqref{rina}],
for $\Lambda=0$ the scalar field becomes ghostlike $(\phi')^2<0$ outside the horizon where $f(r)>0$.
In this case the weak energy condition for the effective energy-momentum tensor 
\eqref{kyoko}
is also violated outside the horizon.
But, because of the overall factor $(1+\Lambda z)$ in \eqref{solp}, 
if $\Lambda<-\frac{1}{z}$, $(\phi')^2$ becomes positive outside the horizon
and then the weak energy condition \eqref{kyoko} is satisfied there. 
Thus, adding a negative cosmological constant can make the black hole solution healthy.

It should also be emphasized that adding a cosmological constant does not modify the 
asymptotic behaviors of the spacetime.
What remains unchanged  even if we add a cosmological constant
is as follows:
\begin{enumerate}
   
\item{The domain of the $r$ coordinate for a given set of $z$ and $K$.}

\item{The asymptotic structure of the spacetime for the domain $0<r<\infty$.}

\item{The metric for the solution of $K=0$, shown in \eqref{hind} and \eqref{mond}.}

\end{enumerate}
The above properties indicate that the asymptotic behaviors of the spacetime 
are determined by the sign of the nonminimal derivative coupling constant $z$ 
and the curvature of the two-dimensional maximally symmetric space $K$,
irrespective of the value of the cosmological constant $\Lambda$.
This is also the unexpected result, from our intuitions in the Einstein gravity,
where the asymptotic behavior in the large $r$ limit 
crucially depends on the value of the cosmological constant. 
For example, if a cosmological constant is sufficiently positive,
we might expect that the spacetime always becomes asymptotically de Sitter
but our analysis has revealed that in our model this is not the case.
For a nonzero $K$, although the cosmological constant $\Lambda$ appears in the 
large $r$ limit, the way of appearance is different from the case of the Einstein gravity
and the coefficients of the leading $O(r^2)$ terms in \eqref{lr1} and \eqref{lr2} are proportional
to $\frac{(1-\Lambda z)^2}{z}$.
Thus for $K=+1$ (and $z>0$) the spacetime is always asymptotically AdS
even if $\Lambda$ is positive.
Similarly, for $K=-1$ (and $z<0$) it is always asymptotically dS even if $\Lambda$ is negative. 
For $K=0$, remarkably, the spacetime structure is also independent of $\Lambda$ and 
purely determined by the sign of $z$.

Moreover, another modification due to the existence of a nonzero cosmological constant
is the appearance of the curvature singularity except for $r=0$ (for a nonzero $K$),
where $g_{rr}$ vanishes.
For example, for $K=+1$ a curvature singularity appears at the finite coordinate position
if the cosmological constant is positive.
Similarly, for $K=-1$ a curvature singularity appears at the finite coordinate position
if the cosmological constant is negative.
In addition,
when the singularity appears at the fiinite coordinate position,
$(\phi')^2<0$ in the large $r$ limit
and there is the violation of the weak energy condition in the large-$r$ limit,
indicating an instability at least at the quantum level.
These properties suggest that in order to obtain the regular spacetime (except for the center at $r=0$),
a choice of a too large positive (negative) cosmological constant is practically forbidden for $K=+1$ ($K=-1$).

%%%%%%%%%%%%%%%%%%%%%%%%%%%%%%%%%%%%%%%%%%%%%%%%%%%%%
\section{Thermodynamic properties of the black hole}

%\subsection{Thermodynamic quantities}

After giving the black hole solutions,
we discuss the thermodynamic properties of our black hole solutions.
In this section, 
we will show the dependence on the reduced Planck mass $m_p$ explicitly.
We consider the black hole solutions of $K=+1$ and $z>0$ given in \eqref{solp}.
We also focus on the case $m_p^2+\Lambda z<0$
for which $(\phi')^2>0$ outside the horizon, as argued in the previous section.
In this section, we will work on the Euclidean frame,
\bea
&&ds_E^2=f(r)d\tau^2+\frac{dr^2}{g(r)}+r^2 d\Omega_{K=+1}^2,
\nonumber\\
&&\phi_E(r)=\phi(r),
\eea
where $\tau=it$ represents the Euclidean time,
and $f(r)$ and $g(r)$ are given in \eqref{solp},
which by giving the $m_p$ dependence back explicitly can be rewritten as
\bea
\label{solp3}
f(r)&=&\frac{1}{12rzm_p}
\Big\{-24m z
+m_p^3r^3\Big(1-\frac{\Lambda z}{m_p^2}\Big)^2
\nonumber\\
&-&3m_p r z\Big(-1+\frac{\Lambda z}{m_p^2}\Big)\Big(3+\frac{\Lambda z}{m_p^2}\Big)
\nonumber\\
&+&
3z^{3/2} \Big(1+\frac{\Lambda z}{m_p^2}\Big)^2{\rm arctan}\big(\frac{m_p r}{\sqrt{z}}\big)
\Big\},
\nonumber\\
g(r)&=&\frac{\big(-2z +r^2(-m_p^2+\Lambda z)\big)^2}
                {4(m_p^2r^2+z)^2f(r)},
\nonumber\\
(\phi_E'(r))^2&=&-\frac{m_p^4 (m_p^2 +\Lambda z)\big(-2r z+r^3 (-m_p^2+\Lambda z)\big)^2}
                 {4(m_p^2r^2+z)^3z f(r)}.
\nonumber \\
\eea
Note that there will be the difference in thermodynamic quantities shown below,
by the overall factor $(8\pi)$ from those in Ref. \cite{rinaldi}.
This difference comes from the definition of the reduced Planck mass $m_p^2=\frac{1}{8\pi G_N}$,
where $G_N$ represents Newton's constant.

Following \cite{rinaldi} (and \cite{hp} for the Schwarzschild-AdS black hole),
we compute thermodynamic quantities of our black hole solutions. 
The temperature of the black hole, $\beta^{-1}$, is given in \eqref{temp_lam},
where $\beta$ represents the periodicity of the $\tau$ coordinate.
Giving the reduced Planck mass $m_p$ back explicitly, it is given by
\bea
\beta&=&\frac{8\pi r_h z}{m_p^2r_h^2+z (2-\Lambda r_h^2)}
=\frac{8\pi m_p x\sqrt{z}}{2m_p^2+x^2 (m_p^2-\Lambda z)},
\eea
where we have introduced the dimensionless horizon position by $x:= m_p r_h/\sqrt{z}$.
For $\Lambda z<-m_p^2$,
the temperature $\beta^{-1}$ is always positive,
decreases for $x<x_{\rm min}$ and increases for $x>x_{\rm min}$,
where $x_{\rm min}:=\frac{\sqrt{2}}{\sqrt{m_p^2-\Lambda z}}m_p$ is the value of $x$ 
for which the temperature $\beta^{-1}$ takes the minimal value.
Note that 
the temperature $\beta^{-1}$ diverges in both the limits of $x\to 0$ and $x\to \infty$,
and hence for a given temperature  
there are always two possible horizon sizes:
one of them is a large black hole $x>x_{\rm min}$ 
and the other is the small black hole $x<x_{\rm min}$,
as for the Schwarzschild-AdS black hole.
In the discussions below,
by ``small'' and  ``large'',
we basically mean that $x<x_{\rm min}$ and $x>x_{\rm min}$, respectively.

%%%%%%%%%%%%%%%%%%%%%%%%%%%%%%
As for the Schwarzschild AdS black hole in the Einstein gravity \cite{hp}, 
the Euclidean action 
\bea
\label{e_action}
S_E&=&
\int d^4x\sqrt{g_E}
{\cal L}_{E},
\nonumber\\
{\cal L}_E
&=&-\frac{1}{2}
\Big\{
m_p^2
\Big(
R_E-2\Lambda
\Big)
\nonumber\\
&-&
\big(g_E^{\mu\nu}-
\frac{z}{m_p^2} G_E^{\mu\nu}\big)
\partial_\mu \phi_E
\partial_\nu \phi_E
\Big\},
\eea
is divergent.
%since our black hole is always asymptotically AdS. 
Hence we need to regularize it
by subtracting the Euclidean action of the background of $m=0$
with the given values of $z$ and $\Lambda$,
 as done in \cite{hp,rinaldi}. 
The regularized Euclidean action is defined by
\bea
\label{reg}
S_{\rm reg}:=S_E-S_{E}^{(0)},
\eea
where
\bea
\label{suki}
S_{E}= 4\pi \beta \int_{r_h}^{\bar r} dr\, r^2\sqrt{f(r) g(r)}{\cal L}_E(r),
\nonumber\\
S_{E}^{(0)}= 4\pi \beta_{(0)} \int_{0}^{\bar r} dr\, r^2\sqrt{f_{(0)}(r) g_{(0)}(r)}{\cal L}^{(0)}_E(r),
\eea
and the quantities with ``$(0)$'' are defined for 
the background of $m=0$, 
$r_h$ is the position of the horizon
and $\bar r$ is the position of the boundary as the regulator of the Euclidean action.
$\beta_{(0)}$ is determined by requiring that  
the periodicity of the $\tau$ direction
and the geometry at the section of $\bar r$ 
in the two backgrounds of $m\neq 0$ and $m=0$
should be identical, namely,
\bea
\label{per}
\beta_{(0)} \sqrt{f_{(0)}(\bar r)}=\beta \sqrt{f(\bar r)}.
\eea
%which determines $\beta_{(0)}$ as the function of $\bar r$.

The explicit computation of Eqs. (\ref{reg}) and \eqref{suki} with the relation \eqref{per}
shows that the divergent parts with positive power of $\bar r$
are canceled and the resultant regularized action contains only the finite value 
\begin{widetext}
\bea
S_{\rm reg}
=
\frac{8\pi^2 x z}{3m_p^2\big(2m_p^2+x^2(m_p^2-\Lambda z)\big)}
\Big[
x
\big(
  m_p^4 (3-2x^2)
+m_p^2(-6+x^2)\Lambda z
+(-3+x^2)\Lambda^2 z^2
\big)
+3(m_p^2+\Lambda z)^2
\arctan(x)
\Big].
\eea
\end{widetext}
For $\Lambda= 0$, we recover the result in \cite{rinaldi}
\bea
S_{\rm reg}
=
\frac{8\pi^2 x z}{3(2+x^2)}
\Big[
 3x-2x^3
+3
\arctan(x)
\Big],
\eea
except for the overall $(8\pi)$ difference 
which arises because of the reason shown in the first paragraph of this section.

The partition function $Z$ is related to the regularized Euclidean action $S_{\rm reg}$
and the Helmholtz free energy $F$ by $\ln Z= - S_{\rm reg}=-\beta F$.
Then we can compute the energy and entropy of the black hole by
$E=\frac{\partial S_{\rm reg}}{\partial \beta}$
and 
$S_{\rm ent}=\beta E -S_{\rm reg}$,
respectively,
which are explicitly written as
\begin{widetext}
\bea
\label{tdm}
E&=&\frac{\pi \sqrt{z}}{3m_p^3 (x^2+1) \big(2m_p^2-x^2(m_p^2-\Lambda z)\big)}
\Big\{
x\Big(
  m_p^6(18-x^2-20x^4-4x^6)
-m_p^4(12+13x^2 -14x^4-6x^6)\Lambda z
\nonumber\\
&-&m_p^2(6+7x^2-8x^4)\Lambda^2 z^2
-x^2(3+2x^2+2x^4)\Lambda^3 z^3
\Big)
\nonumber\\
&+&
3(1+x^2)(m_p^2+\Lambda z)^2
\big(2m_p^2-x^2(m_p^2-\Lambda z)\big)
{\rm arctan}(x)
\Big\},
\nonumber\\
S_{\rm ent}&=&
\frac{8\pi^2 x^2 z}{m_p^2 (1+x^2)\big(2m_p^2-x^2(m_p^2-\Lambda z)\big)}
\Big\{
 m_p^4 (2-x^2-2x^4)
+m_p^2x^2 (-1+x^2)\Lambda z
+x^4\Lambda^2 z^2
\Big\}.
\eea
\end{widetext}
It is straightforward to check that the first law of thermodynamics, $dE=\beta^{-1}dS_{\rm ent}$, is satisfied.
In the limit of $\Lambda=0$,
the above thermodynamic quantities in \eqref{tdm} agree with the results in \cite{rinaldi}
[again except for the overall $(8\pi)$ difference.
Thermodynamic properties of the solutions with $\Lambda=0$ have been investigated in \cite{rinaldi}.
Here  we focus on the case $\Lambda z<-m_p^2$
for which the scalar field is not ghostlike $(\phi')^2>0$ outside the horizon.

First, we discuss the properties of the energy $E$.
Figure 1 shows the region where $E>0$ on the $(x,\frac{\Lambda z}{m_p^2})$ plane,
which corresponds to the region above the drawn curve. 
%%%%%%%%%%%%%%%% Figure I %%%%%%%%%%%%%%%%
\begin{figure*}[ht]
\unitlength=1.1mm
\begin{center}
\begin{picture}(65,40)
  \includegraphics[height=5.0cm,angle=0]{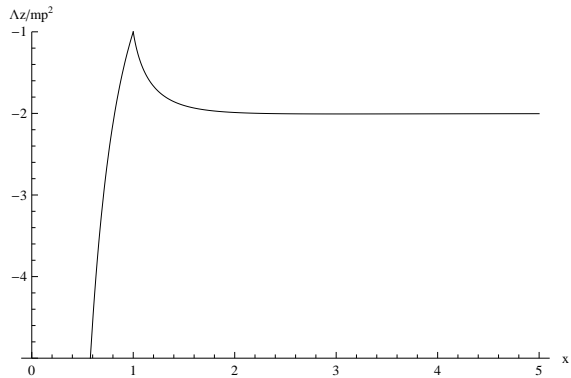}
\end{picture}
\caption{
The region above the curve on the $(x,\frac{\Lambda z}{m_p^2})$ plane 
corresponds to that of the positive energy $E>0$.
A large black hole can have the positive energy only for $-2 m_p^2\leq \Lambda z<-m_p^2$.
 }
  \label{fig:c1}
\end{center}
\end{figure*}
%%%%%%%%%%%%%%%%%%%%%%%%%%%%%%%%%%%%%% 
Note that for $\Lambda z=-m_p^2$, the energy $E$ is always positive.
If $-2m_p^2\leq \Lambda z<-m_p^2$,
for a black hole with the intermediate horizon size $x_{\rm min}<x<x_{1}$,
where $x_1(>x_{\rm min})$ is the value of $x$ for which $E=0$,
the energy $E$ becomes negative.
If $\Lambda z<-2m_p^2$, the energy $E$ becomes negative for a large black hole $x>x_{\rm min}$.
%%%%%%%%%%%%%%%%%%%%%%%%%%%%%%%%%%%%%%%%%%%%%%%%%%

We then discuss the behavior of the entropy $S_{\rm ent}$.
In the small $r_h$ limit ($x\ll 1$), 
\bea
\label{smallr}
S_{\rm ent}= 8\pi ^2 m_p^2 r_h^2-\frac{8\pi^2r_h^4 m_p^2 }{z} (m_p^2+\Lambda z)
%\nonumber\\
%&+&
+ O(r_h^6),
\eea
where the leading order term follows the ordinary area law 
of the black hole entropy $S_{\rm ent}=\frac{A_h}{4G_N}=2\pi m_p^2 A_{h}$
with the horizon area $A_h=4\pi r_h^2$ and $m_p^2=\frac{1}{8\pi G_N}$.
In the large $z$ limit, we also obtain the same leading order behavior as Eq. (\ref{smallr}) following the area law.
On the other hand, 
in the large $r_h$ limit ($x\gg 1$),
\bea
S_{\rm ent}=8\pi^2  (2m_p^2+\Lambda z)r_h^2+ O(r_h^{0}).
\eea
Thus in order for a large black hole to have the positive entropy,
we have to impose $2m_p^2+\Lambda z> 0$.
Fig. 2 shows the region where $S_{\rm ent}>0$ on the $(x,\frac{\Lambda z}{m_p^2})$ plane,
which corresponds to the region above the drawn curve. 
%%%%%%%%%%%%%%%% Figure I %%%%%%%%%%%%%%%%
\begin{figure*}[ht]
\unitlength=1.1mm
\begin{center}
\begin{picture}(65,40)
  \includegraphics[height=5.0cm,angle=0]{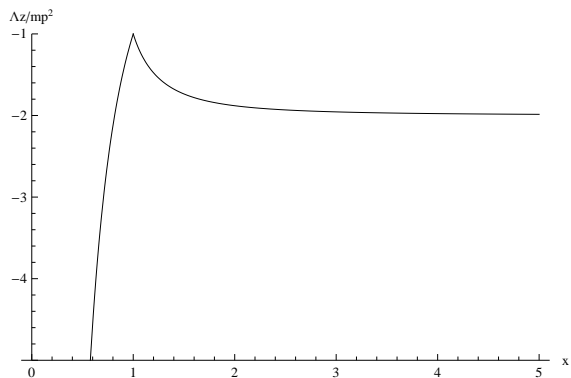}
\end{picture}
\caption{
The region above the curve on the $(x,\frac{\Lambda z}{m_p^2})$ plane 
corresponds to that of the positive entropy $S_{\rm ent}>0$.
A large black hole can have the positive entropy only for $-2 m_p^2\leq \Lambda z<-m_p^2$.
 }
  \label{fig:c1}
\end{center}
\end{figure*}
%%%%%%%%%%%%%%%%%%%%%%%%%%%%%%%%%%%%%% 
Although there is the similarity between Figs. 1 and 2,
the regions of $E>0$ and $S_{\rm ent}>0$ on the $(x,\frac{\Lambda z}{m_p^2})$ plane
do not precisely coincide for $x>1$,
while they coincide for $x<1$.
Note that for $\Lambda z=-m_p^2$, the area law $S_{\rm ent}=\frac{A_h}{4G_N}$
is recovered, as expected from the discussions in Sec. IV A 3.
If $-2m_p^2\leq \Lambda z<-m_p^2$,
only for the intermediate horizon sizes $x_{\rm min}<x<x_{2}$,
where $x_2(>x_{\rm min})$ is the value of $x$ for which $S_{\rm ent}=0$, 
the entropy $S_{\rm ent}$ becomes negative.
If $\Lambda z<-2m_p^2$, the entropy $S_{\rm ent}$ becomes always negative for a large black hole $x>x_{\rm min}$.

Next, the heat capacity $C:=-\beta^{2}\frac{\partial E}{\partial\beta}$
is explicitly given by
\begin{widetext}
\bea
\label{hc}
C&=&\frac{16 \pi^2 x^2 z \big(2m_p^2+x^2(m_p^2-\Lambda z)\big)}
 {m_p^2 (1+x^2)^2\big(2m_p^2-x^2(m_p^2-\Lambda z)\big)^3}
\Big\{
-m_p^6 (4-4x^2-11x^4-4x^6+2x^8)
+m_p^4x^2 (4-2x^2+2x^4+3x^6)\Lambda z
\nonumber\\
&-&m_p^2 x^4(5+4x^2)\Lambda^2 z^2
-x^6(2+x^2)\Lambda^3 z^3
\Big\}.
\eea
\end{widetext}
In the limit of $\Lambda=0$,
the heat capacity \eqref{hc}
again agrees with the result in \cite{rinaldi}.
Let us focus on the case of $\Lambda<0$.
For $\Lambda z=-m_p^2$,
we recover the heat capacity for the Schwarzschild-AdS black hole
\bea
C=\frac{16\pi ^2r_h^2(r_h^2\Lambda-1)}
           {r_h^2\Lambda+1}.
\eea
In Fig. 3, we show the behavior of $C$ (divided by $z$) 
as the function of $x$ for the various choices of $\frac{\Lambda z}{m_p^2}$. 
%%%%%%%%%%%%%%%% Figure II %%%%%%%%%%%%%%%%
\begin{figure*}[ht]
\unitlength=1.1mm
\begin{center}
\begin{picture}(155,40)
  \includegraphics[height=4.5cm,angle=0]{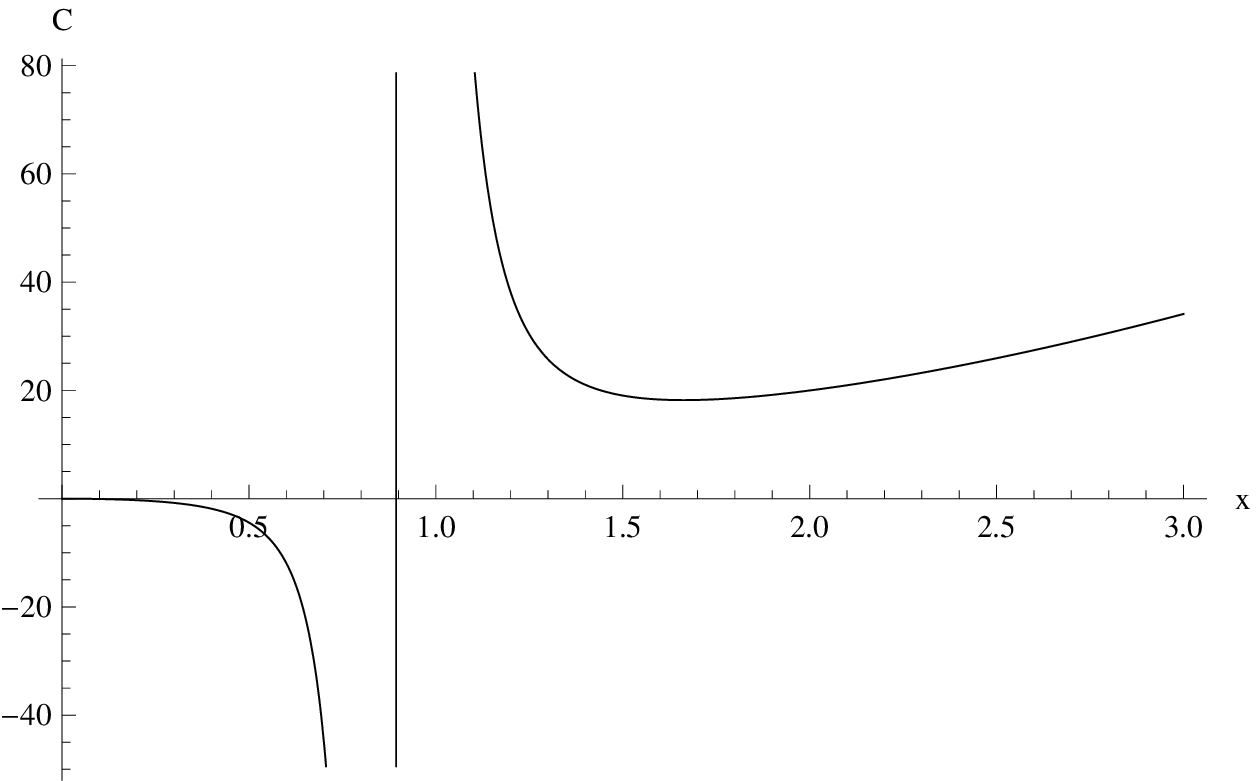}
  \includegraphics[height=4.5cm,angle=0]{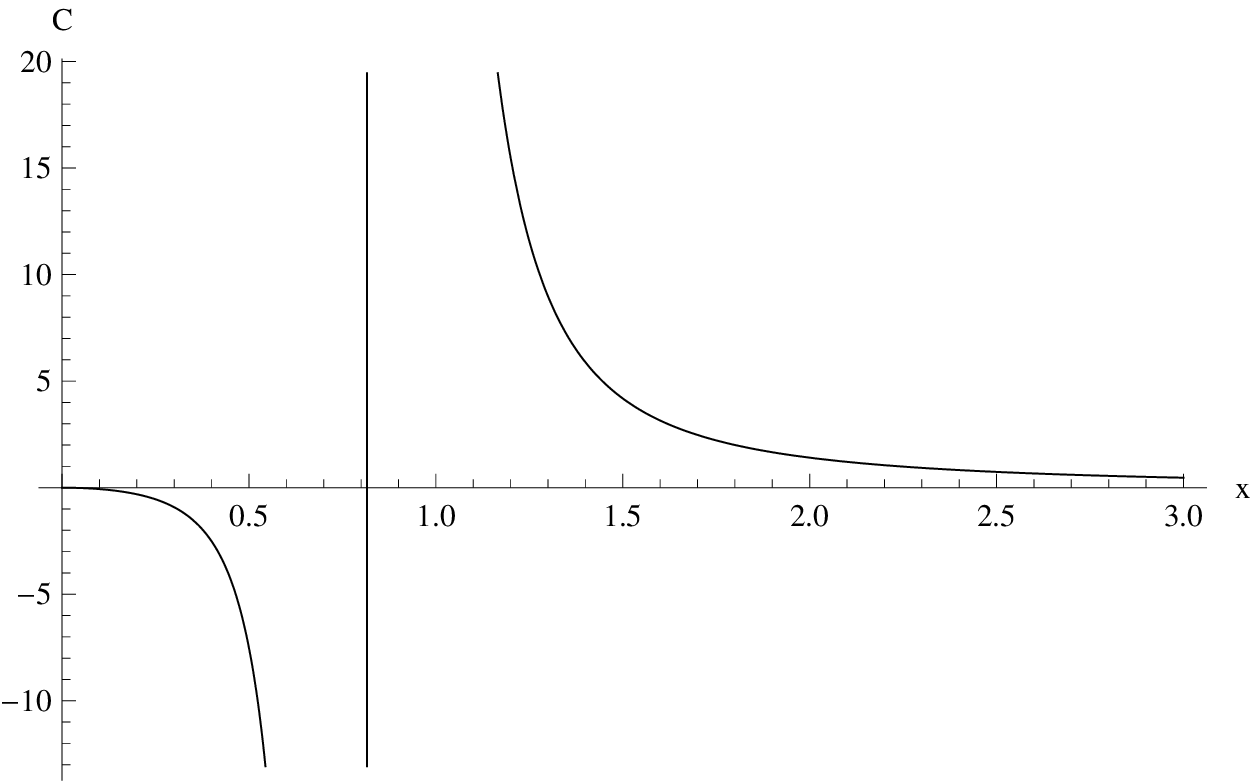}
\end{picture}
\vspace{0.5cm}
\begin{picture}(155,40)
  \includegraphics[height=4.5cm,angle=0]{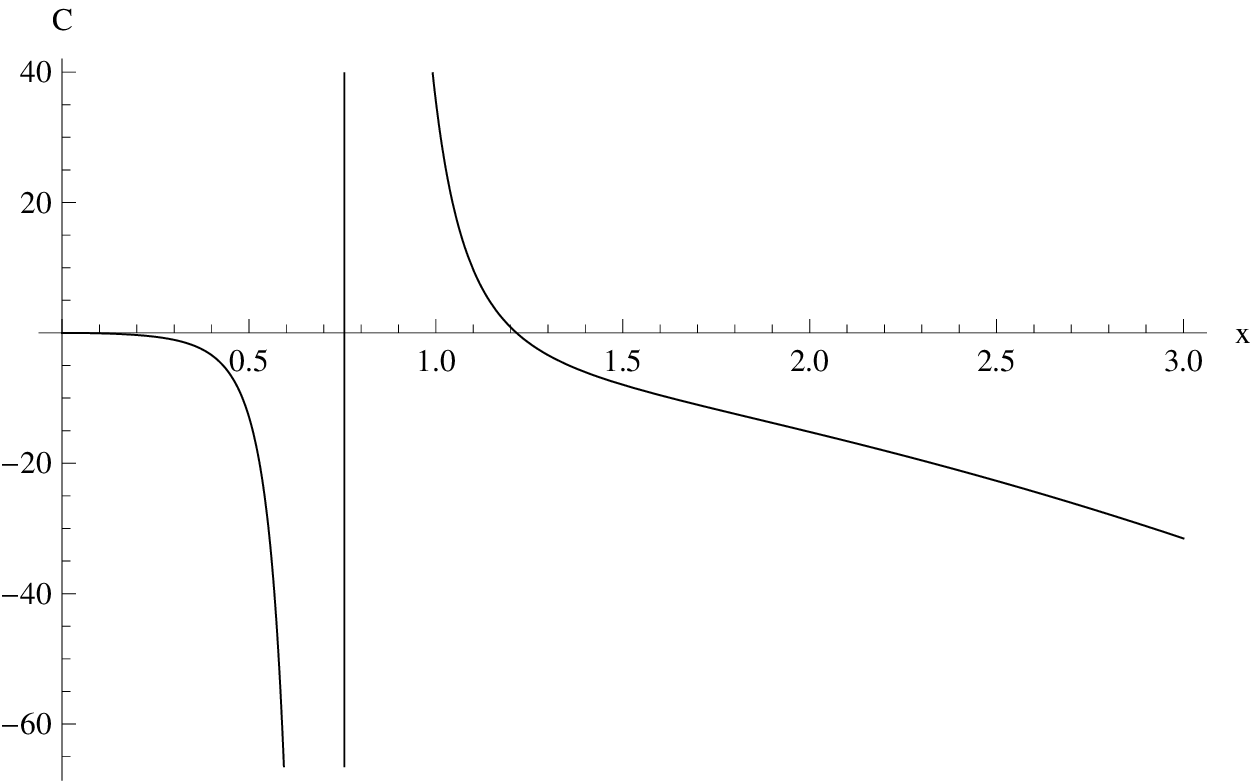}
  \includegraphics[height=4.5cm,angle=0]{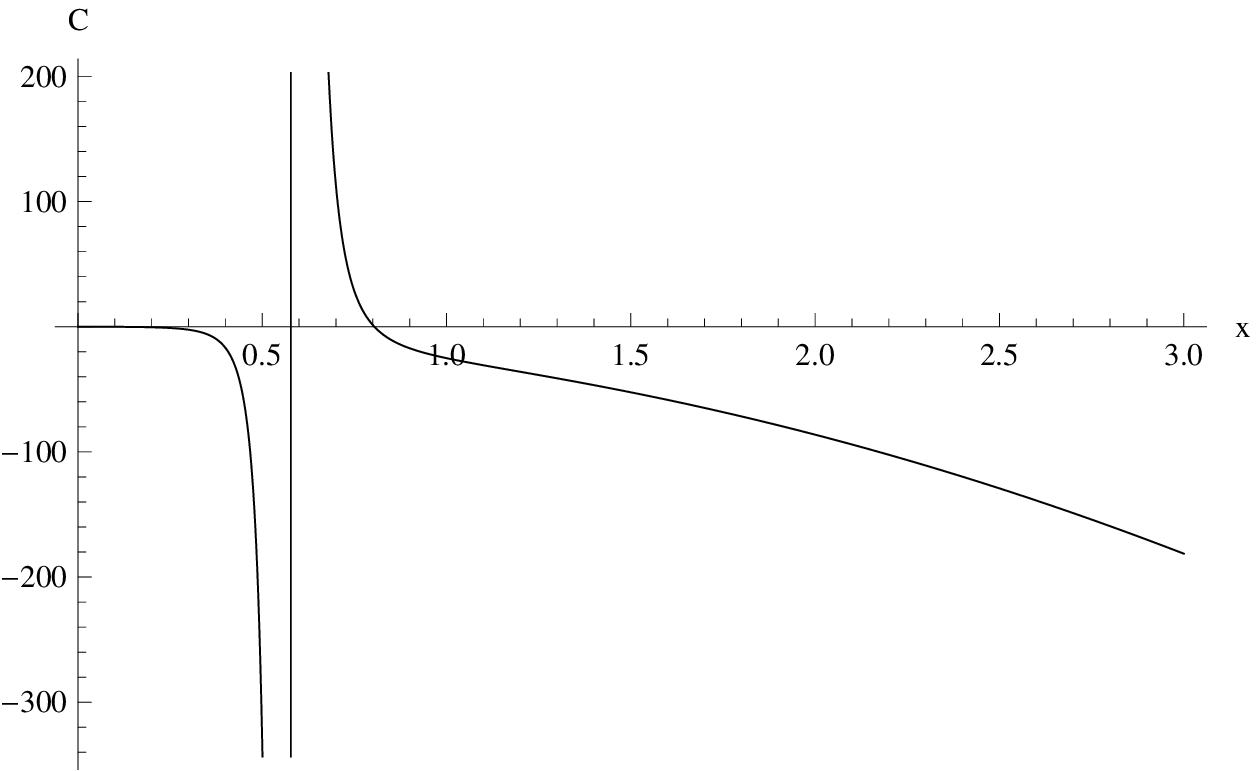}
\end{picture}
\caption{
The heat capacity $C$ (divided by $z$) 
is shown as the function of $x$.
The top-left, top-right, bottom-left and bottom-right panels
show the cases of $\frac{\Lambda z}{m_p^2}=-1.5,-2,-2.5,-5$, respectively.
In all panels, $C$ diverges at $x=x_{\rm min}$.
For $0<x<x $, 
we always have $C<0$.
On the other hand, 
there is the clear difference on the behavior for $x>x_{\rm min}$,
depending on the value of $\Lambda z$.
In the case of $-2m_p^2\leq \Lambda z<-m_p^2$, $C$ is positive for $x>x_{\rm min}$.
On the other hand
in the case of  $\Lambda z<-2m_p^2$, 
$C$ is positive only for the intermediate region $x_{\rm min}<x<x_3$,
where $x_3 (>x_{\rm min})$ is the value of $x$ at which $C=0$.
}
  \label{fig:c1}
\end{center}
\end{figure*} 
%%%%%%%%%%%%%%%% Figure 1%%%%%%%%%%%%%%
The heat capacity \eqref{hc} diverges at $x=x_{\rm min}$ and changes its sign across this point.
There is the clear difference in the behavior of the heat capacity across $\Lambda z=-2m_p^2$.
In Fig. 4 we show the region of $C>0$,
which corresponds to that between two drawn curves.
The left curve in Fig. 4 corresponds to $x=x_{\rm min}$.
%%%%%%%%%%%%%%%% Figure I %%%%%%%%%%%%%%%%
\begin{figure*}[ht]
\unitlength=1.1mm
\begin{center}
\begin{picture}(65,40)
  \includegraphics[height=5.0cm,angle=0]{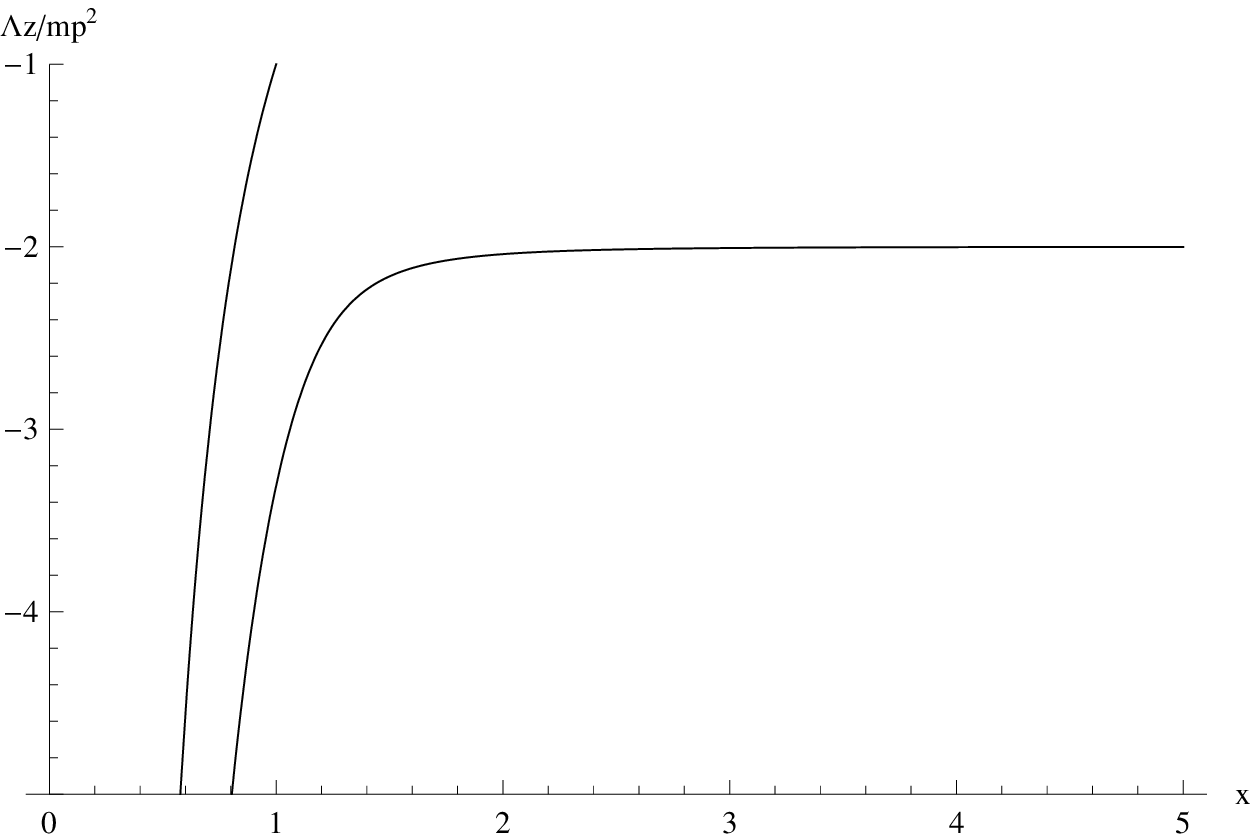}
\end{picture}
\caption{
The region between two curves on the $(x,\frac{\Lambda z}{m_p^2})$ plane
corresponds to that of $C>0$.
The left and right curves show $x=x_{\rm min}$ and $x=x_3$, respectively.
For $-2m_p^2\leq \Lambda z<-m_p^2$, 
a large black hole has a positive heat capacity.
 }
  \label{fig:c1}
\end{center}
\end{figure*}
%%%%%%%%%%%%%%%%%%%%%%%%%%%%%%%%%%%%%% 
If $-2m_p^2\leq \Lambda z<-m_p^2$,
$C>0$ for $x>x_{\rm min}$,
while
$C<0$ for $0<x<x_{\rm min}$,
which is very similar to the case of the Schwarzschild-AdS black hole in the Einstein gravity,
in the sense that a large black hole ($x>x_{\rm min}$) is thermodynamically stable
while a small black hole ($x<x_{\rm min}$) is unstable.
On the other hand,
if $\Lambda z<-2m_p^2$,
$C<0$ for $0<x<x_{\rm min}$ and $x>x_3$,
where $x_3(>x_{\rm min})$ is the value of $x$ at which $C=0$,
while
$C>0$ for $x_{\rm min}<x<x_3$,
indicating that only a black hole with an intermediate horizon size $x_{\rm min}<x<x_{3}$
can be thermodynamically stable.
There is the difference from the case of $\Lambda =0$ discussed in \cite{rinaldi}:
For $\Lambda=0$, there is the region of $x$ where the heat capacity $C$ becomes positive even for $x<x_{\rm min}$,
while in our case the heat capacity is always negative for $x<x_{\rm min}$.

Finally, let us investigate the behavior of the free energy $F=\beta^{-1}S_{\rm reg}$.
In Fig. 5,
the free energy $F$ (divided by $\sqrt{z}$)
is shown as the function of $x$
for the various choices of $\Lambda z<-m_p^2$.
%%%%%%%%%%%%%%%% Figure II %%%%%%%%%%%%%%%%
\begin{figure*}[ht]
\unitlength=1.1mm
\begin{center}
\begin{picture}(155,40)
  \includegraphics[height=4.5cm,angle=0]{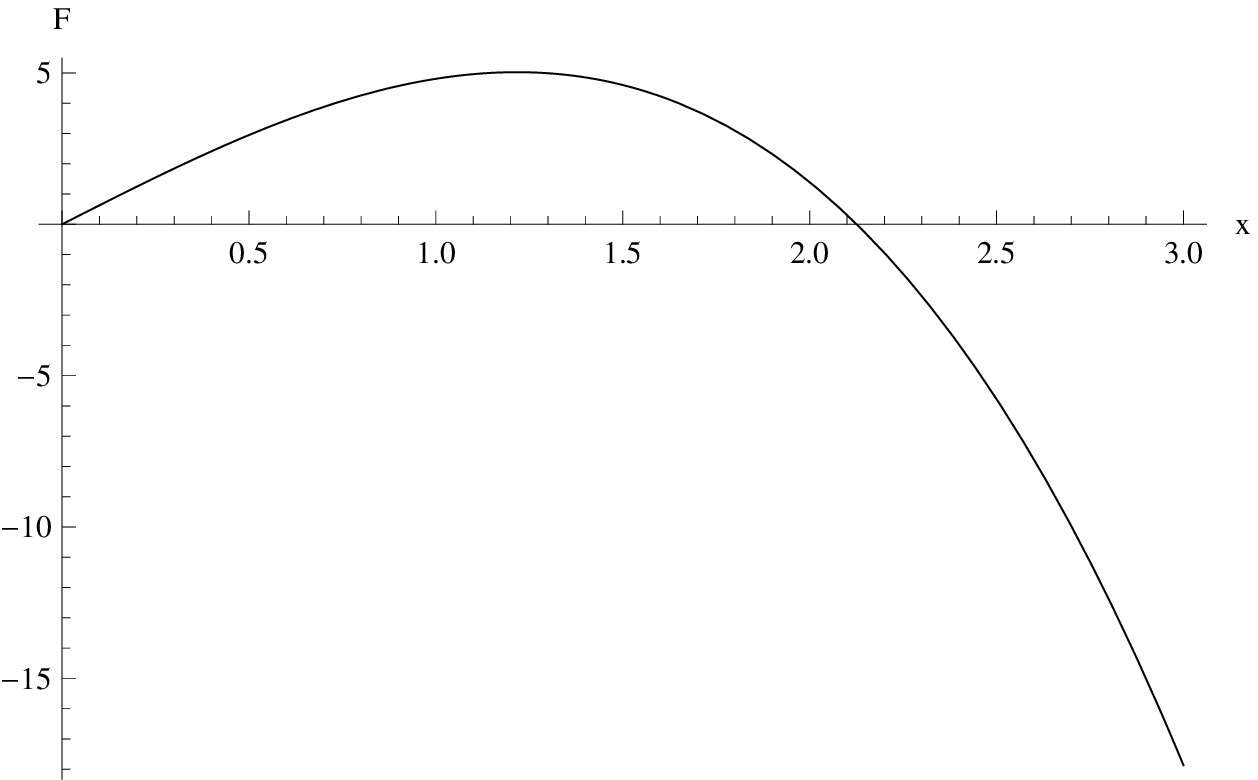}
  \includegraphics[height=4.5cm,angle=0]{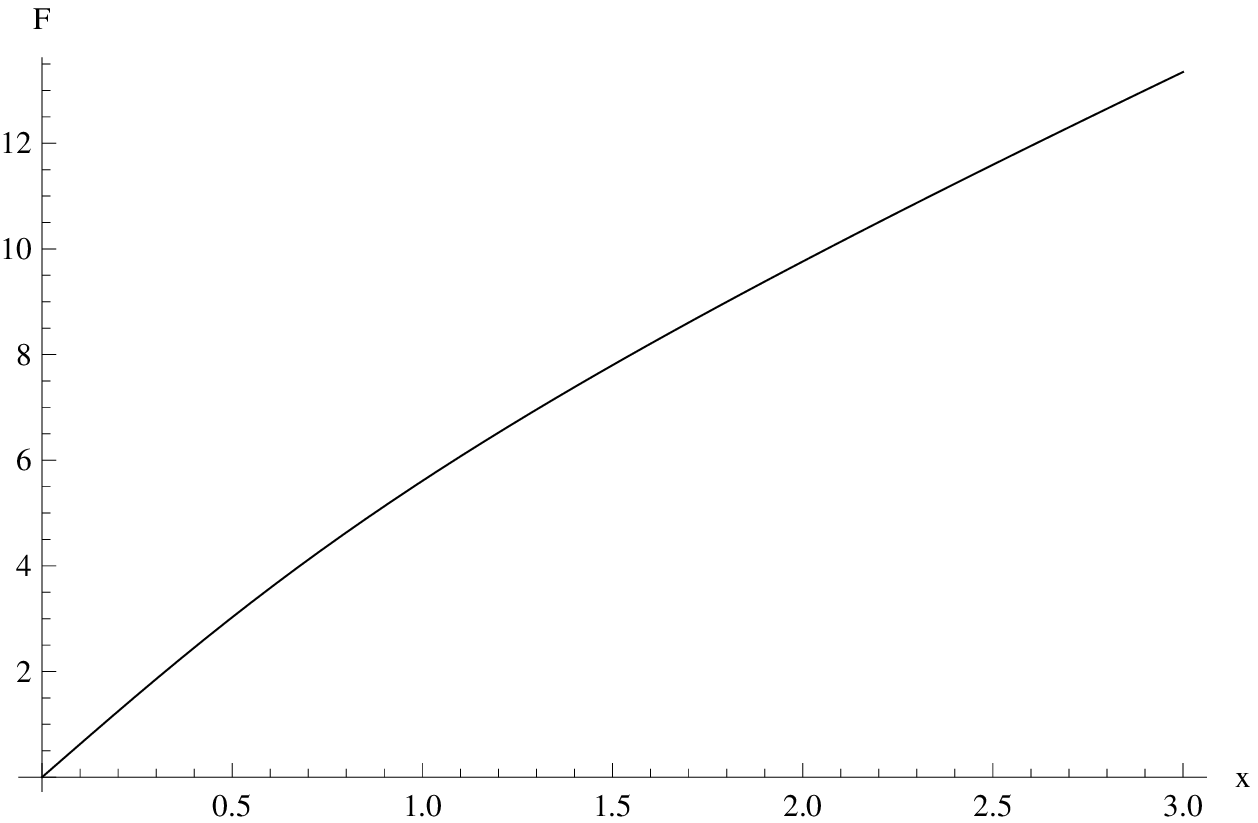}
\end{picture}
\vspace{0.5cm}
\begin{picture}(155,40)
  \includegraphics[height=4.5cm,angle=0]{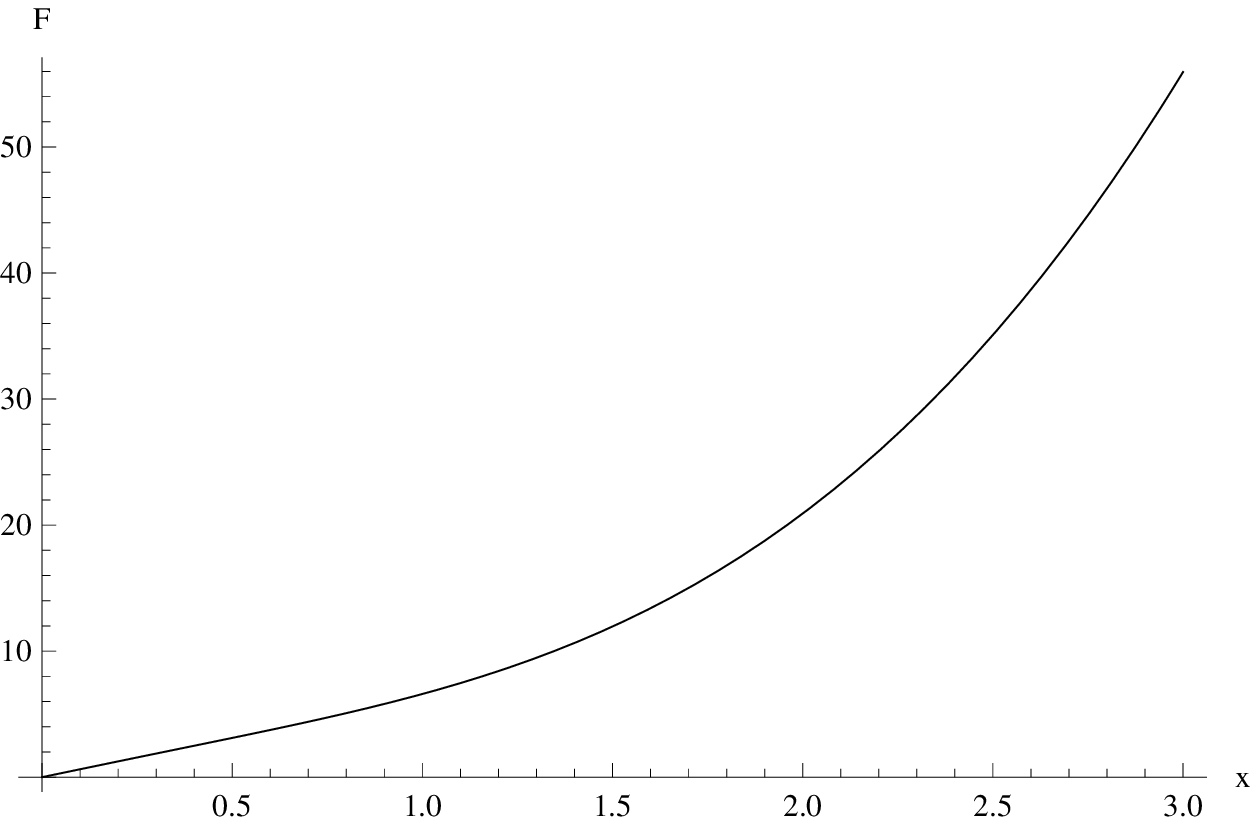}
  \includegraphics[height=4.5cm,angle=0]{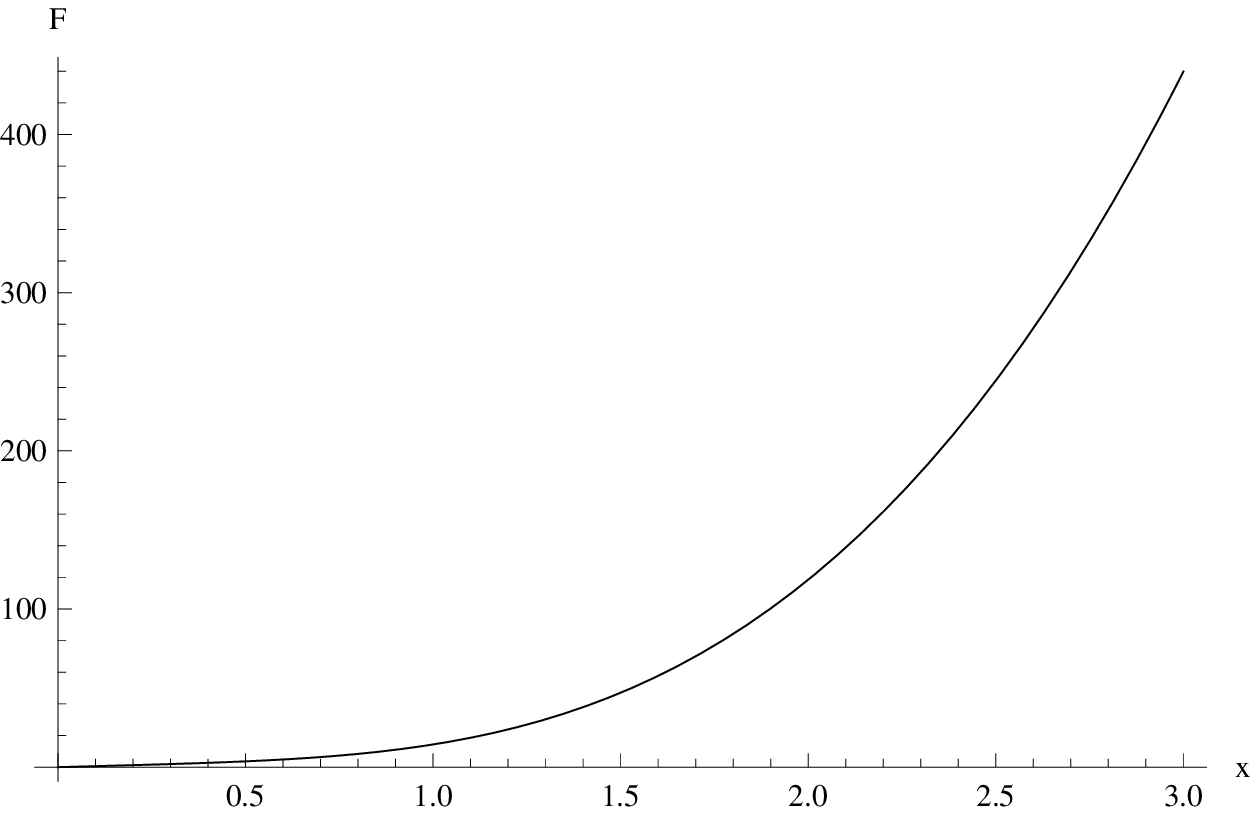}
\end{picture}
\caption{
The free energy $F$ (divided by $\sqrt{z}$)
is shown as the function of $x$.
The top-left, top-right, bottom-left and bottom-right panels
show the cases of $\frac{\Lambda z}{m_p^2}=-1.5,-2,-2.5,-5$, respectively.
}
  \label{fig:c1}
\end{center}
\end{figure*} 
%%%%%%%%%%%%%%%% Figure 1%%%%%%%%%%%%%%
In Fig. 6, the region where the free energy $F>0$ is shown on the $(x,\frac{\Lambda z}{m_p^2})$ plane,
which corresponds to the region below the drawn curve.
%%%%%%%%%%%%%%%% Figure 6 %%%%%%%%%%%%%%%%
\begin{figure*}[ht]
\unitlength=1.1mm
\begin{center}
\begin{picture}(65,40)
  \includegraphics[height=5.0cm,angle=0]{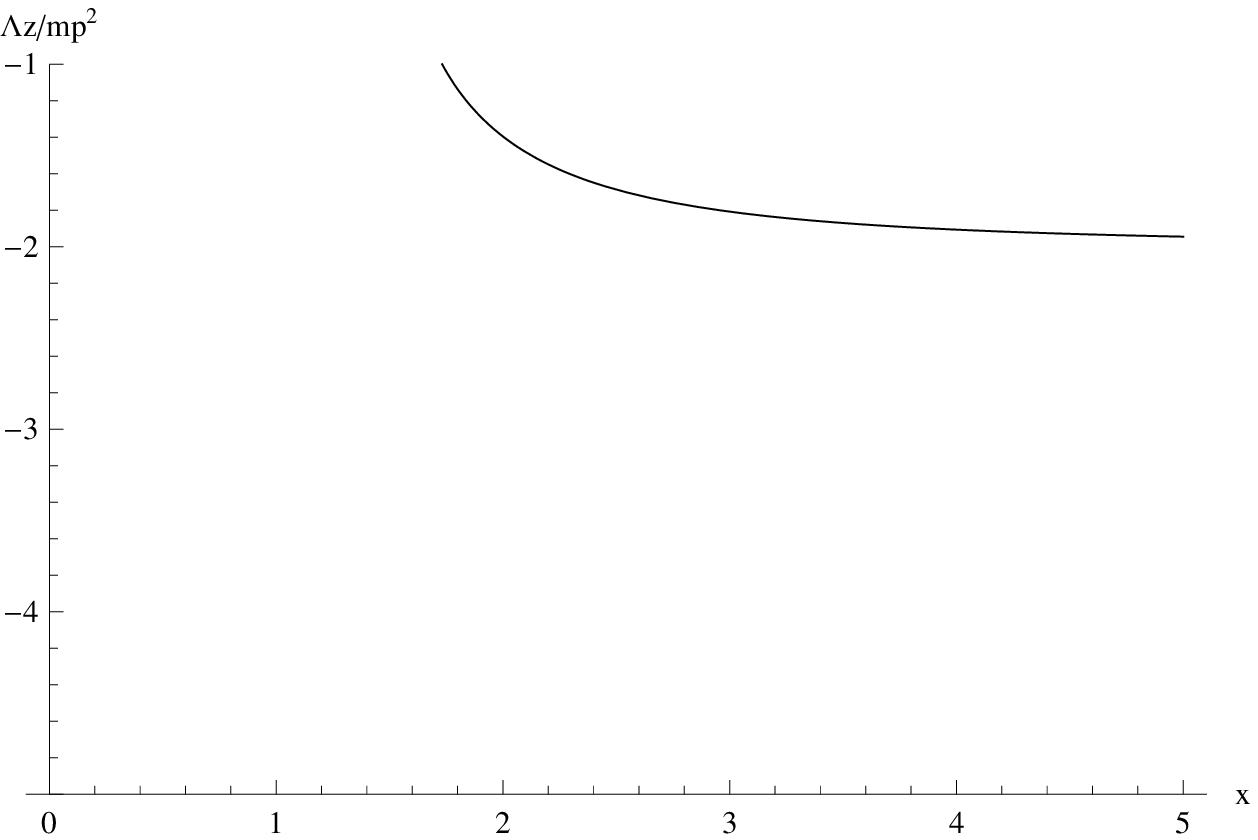}
\end{picture}
\caption{
The region below the curve on the $(x,\frac{\Lambda z}{m_p^2})$ plane
corresponds to the region of the positive free energy $F>0$.
 }
  \label{fig:c1}
\end{center}
\end{figure*}
%%%%%%%%%%%%%%%%%%%%%%%%%%%%%%%%%%%%%% 
As Fig. 5 shows,
the free energy $F$ vanishes at $x=0$ and takes a positive value for a smaller $x$,
irrespective of $\Lambda z$.
There is the clear difference in the behavior of the free energy $F$ for a larger $x$,
depending on whether $\Lambda z$ is greater than $-2m_p^2$ or not.
If $-2m_p^2< \Lambda z<-m_p^2$,
as $x$ increases,
the free energy $F$ increases, but decreases after reaching a maximum, crosses zero and then becomes negative,
which is very similar to the case of the Schwarzschild-AdS black hole.
The free energy takes the maximal value at  
\bea
x&=&x_{\rm peak}
\nonumber\\
&:=&m_p\sqrt{\frac{m_p^2 +\Lambda z-\sqrt{17m_p^4-6m_p^2\Lambda z-7(\Lambda z)^2}}
                                  {2(\Lambda z+2m_p^2)(\Lambda z-m_p^2) }}.
\eea
Thus, a large black hole $x>x_4(>x_{\rm peak})$,
where $x_4$ is the value of $x$ at which $F=0$,
has the negative free energy $F<0$.
If $\Lambda z<-2 m_p^2$,
as $x$ increases,
the free energy $F$ is always positive and increasing.
Thus, in contrast to the above case, the state of the thermal radiation ($x=0$) always has the lowest free energy $F=0$.
Therefore, we expect that for $-2m_p^2<\Lambda z<-m_p^2$ the behavior of thermodynamic quantities is very similar 
to those  of the Schwarzschild-AdS black hole
and the Hawking-Page phase transition \cite{hp} would take place as in the Einstein gravity,
while for $\Lambda z\leq -2m_p^2$ the state of the thermal radiation ($x=0$) always has the minimal free energy
and the Hawking-Page phase transition would not take place.
Note that the behavior of the free energy $F$ and the other thermodynamic quantities
for $\Lambda z=-m_p^2$ is precisely the same as in the case of the Schwarzschild-AdS black hole,
and the Hawking-Page phase transition can take place \cite{hp}.
%According to the AdS/CFT correspondence,
%the Hawking-Page phase transition in the AdS spacetime
%corresponds to the transition between the confinement-deconfinement phases in
%the dual conformal field theory 
Our results may be also useful to obtain the holographic interpretation of the scalar-tensor theory
with nonminimal derivative coupling to the Einstein tensor, along the line of \cite{witten}. 

%%%%%%%%%%%%%%%%%%%%%%%%%%%%%%%%%%%%%%%%%%%%%%%%%%%%%%%%%
\section{Solutions in the five-dimensional model}

Before closing, we present the corresponding solutions in the five-dimensional version of the theory \eqref{action}.
The metric ansatz is the same as \eqref{ansatz}
except that $d\Omega_K^2$ denotes the three-dimensional maximally symmetric space.
In this section, we again set $m_p=1$.

The main features of the solutions are the same as 
those of the corresponding solutions of the four-dimensional model.
The extension to the higher-dimensional case
is also straightforward,
although we do not explicitly present the solutions of 
the models with more than six spacetime dimensions.

\subsection{The solutions for $K=+1$}

\subsubsection{$z\neq  -\frac{1}{\Lambda}$}

The solution for $K=+1$ is given by

\bea
f(r)&=&\frac{1}{24r^2z}
\Big(-48mz +r^4(1-\Lambda z)^2
\nonumber\\
&-&6r^2 z(-1+\Lambda z)(3+\Lambda z)
\nonumber\\
&+&18z^2 (1+\Lambda z)^2
\ln\Big(\frac{r^2}{3z}+1\Big)
\Big),
\nonumber\\
g(r)&=&\frac{\big(-6z +r^2(-1+\Lambda z)\big)^2}
                {4(r^2+3z)^2f(r)},
\nonumber\\
(\phi'(r))^2&=&-\frac{(1+\Lambda z)\big(-6r z+r^3 (-1+\Lambda z)\big)^2}
                 {4(r^2+3z)^3z f(r)},
\eea
where the domain of the coordinate $r$
is given by
$0<r<\infty$ for $z>0$ and by $0<r<\sqrt{-3z}$ for $z<0$.
The overall normalization of $f(r)$
is chosen to recover the asymptotic structure
of the Schwarzschild-(A)dS
solution in the small $r$ limit.
We obtain
\bea
\label{srpc3}
f(r)= -\frac{2m}{r^2}+1-\frac{\Lambda r^2}{6}+\frac{(1+\Lambda z)^2r^4}{108z^2}+O(r^6),
%+O(r^4),
\eea
and $g(r)\approx 1/f(r)$.
The explicit dependence on the coupling constant $z$ appears in the $O(r^4)$ term,
by which we could distinguish the present model from the Einstein gravity with a cosmological constant. 
Note that the $z$-dependent correction vanishes in the limit $z\to -\frac{1}{\Lambda}$
which corresponds to the case discussed in Sec. VI A 2. 
In the large $|z|$ limit, we obtain
\bea
\label{largez3}
f(r)&=& 1-\frac{2m}{r}-\frac{\Lambda }{6}r^2 +\frac{\Lambda^2}{108}r^4
+O(\frac{1}{|z|}),
\nonumber\\
f(r)g(r)&=&\Big(1-\frac{r^2\Lambda}{6}\Big)^2+O(\frac{1}{|z|}).
\eea
For $z>0$,
in the large $r$ limit,
the effects of the nonminimal derivative coupling become more important and 
we obtain
\bea
f(r)&=&
\frac{r^2 (\Lambda  z-1)^2}{24z}-\frac{1}{4} (\Lambda  z-1) (\Lambda  z+3)
\nonumber\\
&+&O(r^{-2}).
\eea
Thus the asymptotic structure in the large $r$ limit is 
AdS spacetime
with the effective cosmological constant $-\frac{(1-\Lambda z)^2}{4z} (<0)$.
As for the four-dimensional solution,
for $z>0$, 
$(\phi'(r))^2$ can be positive outside the horizon for $\Lambda <-\frac{1}{z}$
and the scalar field does not become ghostlike,
because $f(r)>0$ for $r>r_h$,
where $r_h$ is the position of the horizon so that $f(r_h)=0$.

The derivative of the scalar field with respect to the proper length, $\frac{1}{\sqrt{|g|}}|\phi'(r)|$,
remains finite at the horizon.
As discussed in the previous section,
the temperature of the black hole $\beta^{-1}$ is given by 
\bea
\beta=\frac{12\pi z r_h }{r_h^2\big(1-\Lambda z\big)+ 6z}.
\eea
For a large $z$ and fixed $\Lambda z$,
the temperature of the five-dimensional Schwarzschild black hole 
$\beta= 2\pi r_h$ is recovered.
On the other hand, for both large $z$ and $(-\Lambda) z$ ($\Lambda<0$),
$ \beta=\frac{2\pi  r_h }{1-\frac{\Lambda}{6}r_h^2}$,
which slightly does not agree with the temperature of the five-dimensional
Schwarzschild-AdS black hole
$ \beta=\frac{2\pi  r_h }{1-\frac{\Lambda}{3}r_h^2}$,
because \eqref{largez3} is not precisely the same
as the metric of the Schwarzschild-AdS.

For $z>0$,
the point of $g(r)=0$,
$r=\sqrt{\frac{6z}{\Lambda z-1}}$,
becomes a curvature singularity
because the invariant $R^{\alpha\beta\mu\nu}R_{\alpha\beta\mu\nu}$ diverges there.
This point appears at a finite coordinate position  for $\Lambda>\frac{1}{z}$,
while for $\Lambda<\frac{1}{z}$ there is no such singularity.
%This singularity is hidden by the horizon.

For $z<0$ and $\Lambda>\frac{1}{(-z)}$,
a singularity appears at $r=\sqrt{\frac{6(-z)}{1+\Lambda (-z)}}(<\sqrt{-3z})$
other than $r=0$.
%This singularity is not hidden by the horizon.
%because $r_h<\sqrt{\frac{6(-z)}{1+\Lambda (-z)}}$.

%The other important observation for $z>0$
%is that 
%because $f(r)>0$ for $r>r_h$
%for $\Lambda z<-1$, $(\phi')^2>0$ outside the horizon,
%while for $\Lambda z>-1$ including the case of $z=0$, $(\phi')^2<0$ outside the horizon.
%Thus adding the negative cosmological constant ensures the positively of $(\phi')^2$ outside the horizon.

\subsubsection{$z= -\frac{1}{\Lambda}$}

For $z=-\frac{1}{\Lambda}$, 
the solution is given by
\bea
f(r)&=&-1-\frac{2m}{r^2}-\frac{\Lambda}{6}r^2,\quad
g(r)=\frac{1}{f(r)},
\nonumber\\
\phi'(r)&=&0.
\eea
Thus the scalar field becomes trivial.

\subsection{The solutions for $K=-1$}

\subsubsection{$z\neq -\frac{1}{\Lambda}$}

The solution for $K=-1$ is given by
\bea
f(r)&=&\frac{1}{24r^2 z}
\Big(-48mz +r^4(1-\Lambda z)^2
\nonumber\\
&+&6r^2 z(-1+\Lambda z)(3+\Lambda z)
\nonumber\\
&+&18z^2 (1+\Lambda z)^2
\ln\Big(-\frac{r^2}{3z}+1\Big)
\Big),
\nonumber\\
g(r)&=&\frac{\big(6z +r^2(-1+\Lambda z)\big)^2}
                {4(r^2-3z)^2f(r)},
\nonumber\\
(\phi'(r))^2&=&-\frac{(1+\Lambda z)\big(6r z+r^3 (-1+\Lambda z)\big)^2}
                 {4(r^2-3z)^3z f(r)},
\eea
where the domain of the coordinate $r$ is given by
$0<r<\infty$ for $z<0$ and by $0<r<\sqrt{3z}$ for $z>0$.
The overall normalization of $f(r)$
is chosen to recover 
the solution in the Einstein gravity
in the small $r$ limit.
We obtain
\bea
\label{srpc}
f(r)= -\frac{2m}{r^2}-1-\frac{\Lambda r^2}{6}-\frac{(1+\Lambda z)^2r^4}{108z^2}+O(r^6),
%+O(r^4),
\eea
and $g(r)\approx 1/f(r)$.
The explicit dependence on the coupling constant $z$ appears in the $O(r^4)$ term,
by which we could distinguish the present model from the Einstein gravity with a cosmological constant. 
Note that the $z$-dependent correction vanishes in the limit $z\to -\frac{1}{\Lambda}$
which corresponds to the case discussed in Sec. VI B 2.
In the large $|z|$ limit, we obtain
\bea
\label{largez4}
f(r)&=& -1-\frac{2m}{r}-\frac{\Lambda }{6}r^2 -\frac{\Lambda^2}{108}r^4
+O(\frac{1}{|z|}),
\nonumber\\
f(r)g(r)&=&\Big(1+\frac{r^2\Lambda}{6}\Big)^2+O(\frac{1}{|z|}).
\eea
For $z<0$,
in the large $r$ limit,
the effects of the nonminimal derivative coupling become more important and 
we obtain
\bea
f(r)&=&
\frac{r^2 (\Lambda  z-1)^2}{24 z}+\frac{1}{4} (\Lambda  z-1) (\Lambda  z+3)
\nonumber\\
&+&O(r^{-2}).
\eea
Thus
the asymptotic structure in the large $r$ limit is dS spacetime
with the effective cosmological constant $\frac{(1-\Lambda z)^2}{4(-z)} (>0)$.

The derivative of the scalar field with respect to the proper length, $\frac{1}{\sqrt{|g|}}|\phi'(r)|$,
remains finite at the horizon.
For $z<0$,
the point of $g(r)=0$,
$r=\sqrt{\frac{6(-z)}{(-\Lambda)(-z)-1}}$,
becomes a curvature singularity
because the invariant  $R^{\alpha\beta\mu\nu}R_{\alpha\beta\mu\nu}$ diverges there.
This point appears for $\Lambda<\frac{1}{z}$,
while for $\Lambda>\frac{1}{z}$ there is no such singularity.
For $z>0$ and $\Lambda<-\frac{1}{z}$,
a singularity appears at $r=\sqrt{\frac{6z}{1-\Lambda z}}(<\sqrt{3z})$
other than $r=0$.
As for the four-dimensional solution,
for $z<0$, 
$(\phi'(r))^2$ can be positive in the large-$r$ region for $\Lambda >\frac{1}{(-z)}$
and the scalar field does not become ghostlike,
because of $f(r)<0$.

\subsubsection{$z= -\frac{1}{\Lambda}$}

For $z=-\frac{1}{\Lambda}$, 
the solution is given by
\bea
f(r)&=&-1-\frac{2m}{r^2}-\frac{\Lambda}{6}r^2,\quad
g(r)=\frac{1}{f(r)},
\nonumber\\
\phi'(r)&=&0.
\eea
Thus the scalar field becomes trivial.

\subsection{The solutions for $K=0$}

The solution for $K=0$ is given by
\bea
\label{mond2}
f(r)&=&-\frac{2m}{r^2} +\frac{r^2}{24z},
\quad
g(r)=\frac{1}{4f(r)},
\nonumber\\
(\phi'(r))^2&=&-\frac{1+\Lambda z}{4z f(r)}.
\eea
This solution is singular only at $r=0$.
The scalar field becomes trivial for $z=-\frac{1}{\Lambda}$.

\subsection{Summary}

%%%%%%%%%%%%%%%%%%%%%%%%%%%%%%%%%%%%%%%%%%%%%%%%%%%%%
In Table III,
the properties of the five-dimensional solutions
are summarized.
The essential properties remain the same as those in the four-dimensional ones.
``Domain'', ``Singularities'' and ``Asymptotic behavior'' correspond to 
the domain of the $r$ coordinate, the position of the curvature singularities
and the asymptotic behavior of spacetime in the $r\to \infty$ limit,  respectively.
%\newpage
%%%%%%%%%%%%%%%%%%%%%%%%%%%%%%%%%%%%%%%%%%%%%%%%%%%%
\begin{widetext}
\begin{table}
\begin{minipage}[t]{.80\textwidth}
\caption{
The properties of solutions in the five-dimensional model
}
\begin{center}
\label{table3}
\begin{tabular}{|c|c|c|c|c|c|c|c|c|
}
\hline
&\multicolumn{2}{c|}{$K=+1$}
&\multicolumn{2}{c|}{$K=-1$}
&\multicolumn{2}{c|}{$K=0$}
\\ \hline  
 & $z>0$ 
 & $z<0$
 & $z>0$ 
 & $z<0$ 
 & $z>0$ 
 & $z<0$
\\ \hline\hline
Domain
& $0<r<\infty$
& $0<r<\sqrt{-3z}$
& $0<r<\sqrt{3z}$
& $0<r<\infty$
& $0<r<\infty$
& $0<r<\infty$ 
  \\ \hline
 Singularities
& $r=0$
& $r=0$
& $r=0$
& $r=0$
& $r=0$
& $r=0$
 \\ 
& $r=\sqrt{\frac{6z}{\Lambda z-1}}$
& $r=\sqrt{\frac{6(-z)}{1+\Lambda (-z)}}$
& $r=\sqrt{\frac{6z}{1-\Lambda z}}$
& $r=\sqrt{\frac{6(-z)}{\Lambda z-1}}$
& 
& 
\\ 
&  ($\Lambda>\frac{1}{z}$)
&   ($\Lambda>\frac{1}{(-z)}$)
&  ($\Lambda<-\frac{1}{z}$)
&  ($\Lambda<-\frac{1}{(-z)}$)
& 
& 
\\
\hline
Asymptotic behavior
& AdS 
& -
& -
& dS
& AdS 
& dS
 \\ \hline
\end{tabular}
\end{center}
\end{minipage}
\end{table}
\end{widetext}
%%%%%%%%%%%%%%%%%%%%%%%%%%%%%%%%%%%%%%%%%%%%%%%%%%%%%
%%%%%%%%%%%%%%%%%%%%%%%%

\section{Conclusions}

We have obtained the black hole type solutions 
in the scalar-tensor theory with nonminimal derivative coupling
to the Einstein tensor
whose action is given by \eqref{action}.
Although we have discussed both the four- and five-dimensional models,
here we mainly summarize the properties of the four-dimensional solutions
because the behaviors of the spacetime and scalar field
 are not sensitive to the spacetime dimensionality. 
Adding a cosmological constant modifies
the features of the solutions in some degree.
The effect of the nonminimally derivative coupling 
becomes more important in the large $r$ regions,
while the solutions 
are approximately the same as those in the Einstein gravity 
in the small $r$ regions,
where the coordinate $r$ given in \eqref{ansatz}
becomes either timelike or spacetime,
 depending on the sign of $f(r)$ and $g(r)$.
For the special choice of nonminimal coupling constant $z=-\frac{1}{\Lambda}$, 
only the solutions with trivial scalar field configurations $\phi={\rm const}$
were obtained
and the solutions in the Einstein gravity 
with the same cosmological constant
were reproduced.
%Adding the negative cosmological constant 
%could also ensure the positivity of $(\phi')^2$ outside the horizon. 

For the other choices of the coupling constant, $z\neq -\frac{1}{\Lambda}$,
more nontrivial solutions were obtained.
The asymptotic structure of the spacetime depends 
on the sign of $z$ and $K$.
For a two-sphere ($K=+1$),
for a positive coupling constant ($z>0$)
with a cosmological constant smaller than the inverse of the coupling constant
($\Lambda<\frac{1}{z}$)
the spacetime is regular except for $r=0$,
while 
for $\Lambda>\frac{1}{z}$
the spacetime also becomes singular at a finite coordinate position.
For a two-hyperboloid ($K=-1$),
for a negative coupling constant ($z<0$)
with a cosmological constant larger than the inverse of the coupling constant
($\Lambda>-\frac{1}{(-z)}$)
the spacetime is regular except for $r=0$,
while 
for $\Lambda<-\frac{1}{(-z)}$
the spacetime also becomes singular at a finite coordinate position.
For a two-dimensional flat space ($K=0$),
we recover the Bianchi I universe
approaching the dS spacetime.
For all the above cases, the domain of the coordinate $r$
is given by $0<r<\infty$.
On the other hand,
in the cases of $K>0$ with $z<0$ and $K<0$ with $z>0$,
the spacetime is bounded at a finite $r$ given by the coupling constant,
but the proper distance to the boundary becomes infinite.
We have also found that in the case of the ordinary black hole spacetime for $K=+1$ and $z>0$,
$(\phi'(r))^2$ can be positive outside the horizon 
and the weak energy condition for the effective energy-momentum tensor 
of the scalar field
can be satisfied only for $\Lambda<-\frac{1}{z}$.
This indicates the importance of adding a nonzero cosmological constant in our model 
to get the black hole solutions which are healthy outside the horizon.
Note that for the same values of the cosmological constant
no curvature singularity is formed except for that at the center $r=0$.

We then summarize the asymptotic properties of the solutions
where the domain of the coordinate $r$ is given by $0<r<\infty$.  
In the large $r$ region, 
the effective cosmological constant is proportional to  $-\frac{1}{4z}$.
Thus,
irrespective of the value of the cosmological constant $\Lambda$,
for a two-sphere ($K=+1$) the effective cosmological constant is negative and 
hence the spacetime approaches AdS,
while for a two-hyperboloid ($K=-1$),
the effective cosmological constant is positive and 
hence the spacetime approaches dS.
For a two-dimensional flat space ($K=0$),
the metric form does not explicitly depend on $\Lambda$,
while it appears in the amplitude of the scalar field. 
In all the cases,
the spacetime is not asymptotically flat for a nonzero $z$.

Thermodynamic properties of our black hole solutions have also been investigated.
We have computed the energy, entropy, heat capacity and Helmholtz free energy of the black holes,
along the same way for the case of the Schwarzschild-AdS black hole.
The expressions of these thermodynamic quantities
have generalized the results obtained without a cosmological constant $\Lambda=0$
\cite{rinaldi} to the case with a cosmological constant.
We have found that 
the entropy for a large black hole with $r_h\gg\sqrt{z}$ (setting $m_p=1$),
where $r_h$ represents the horizon radius,
can become positive only for $\Lambda z\geq  -2$,
while the leading behavior of the entropy for the small black hole $r_h\ll\sqrt{z}$
follows the ordinary area law of the black hole entropy. 
For $\Lambda z=-1$,
we have completely recovered thermodynamic properties of the Schwarzschild-AdS black holes.
For $-2\leq \Lambda z<-1$,
thermodynamic quantities behave in the similar way as that of the Schwarzschild-AdS black hole.
For $\Lambda z<-2$,
the behavior of thermodynamic quantities
is different from those of the Schwarzschild-AdS black hole and also of the case of $\Lambda=0$,
for example, the heat capacity is positive only for the black hole with an intermediate size.

Before closing this paper, we would like to mention the subjects related to our solutions,
especially focusing on the case of the black hole solutions ($K=+1$).
The asymptotically AdS property of the five-dimensional black hole solution
would be useful for constructing the Randall-Sundrum-type cosmological braneworld models \cite{rs,rs3}
in the given class of the scalar-tensor theory.
The modification of the cosmological brane dynamics may appear in the following two ways:
One is the modification of the bulk spacetime metric, as explicitly investigated in this paper.
The other is the modification of the junction conditions 
because of the existence of the nonminimal derivative coupling to the Einstein tensor
(see \cite{padila} for the junction conditions in the generalized Galileon scalar-tensor theory).
The effective four-dimensional cosmological dynamics on the brane
 may involve the energy exchange process between the ordinary matter localized on the brane and the scalar field.
On the other hand, the investigation of the stability and the holographic interpretation of our 
asymptotically AdS black hole solutions  
would also tell us the other essential aspects of 
the scalar-tensor theory with nonminimal derivative coupling to the Einstein tensor.

Another important issue is whether it is possible for us to obtain the asymptotically flat black hole solutions
in the theory \eqref{action},
which may be useful for testing our model in astrophysical environments.
At least,  it is quite difficult to obtain the analytic asymptotically flat black hole solutions in the present model.
In integrating the scalar field equation of motion \eqref{kir} 
and obtaining \eqref{sca2} in the four-dimensional case,
we set the integration constant to be zero,
which is equivalent to assume $G_{rr}=\frac{1}{z}g_{rr}$.
The situation in the five-dimensional case is also similar to that in the four-dimensional case.
Thus for such a choice it is impossible to obtain the asymptotically flat solution
where all components of the Einstein tensor approach zero at the asymptotic infinity.
One approach to the asymptotically flat solution
might be to allow for a nonzero integration constant in integrating \eqref{kir}
but the explicit integration of the remaining equations of motion would be too involved analytically.
If we assume that both $\phi$ and $\phi'$ are regular at the horizon,
it is possible to see
that for a nonzero cosmological constant
a new asymptotically flat black hole solution cannot be obtained.
Integrating the scalar field equation of motion \eqref{kir}, we find
\bea
\label{cw}
\frac{f^{\frac{1}{2}}}{g^{\frac{3}{2}}}
\Big(
\frac{rf'}{f}
-\big(Kg-1\big)
-\frac{r^2 g}{z}
\Big)
\phi'
=D,
\eea
where $D$ is an integration constant.
If we assume the existence of a new black hole solution in which $\phi'$ is regular at the horizon
and expect that in the near horizon limit it behaves as
$f\sim 0$, $g\sim \frac{1}{f}$ and $f'\sim {\rm const}$,
the left-hand side of \eqref{cw} at the horizon vanishes, leading to $D=0$.
Therefore, within the assumption of the regular $\phi'$
it does not allow for the solution with $D\neq 0$ \cite{rinaldi}.
Moreover, $D=0$ gives the conclusion that the left-hand side of \eqref{cw}
must vanish for any $r$ and hence only the solution must be of $\phi'=0$
(unless $G_{rr}=\frac{1}{z}g_{rr}$),
which gives only the Schwarzschild solution for the asymptotically flat case.
It may be interesting to look for the solutions
by replacing the cosmological constant $\Lambda$ in \eqref{action} with a nonconstant potential term $V(\phi)$,
where the scalar field equation of motion \eqref{sca} receives the correction given by $V'(\phi)$,
although it is beyond the present model.

%%%%%% Structure formation %%%%%%%%%%%%%%
In order to discriminate the scalar-tensor theory with nonminimal derivative coupling to the Einstein tensor
from the various scalar-tensor theories other than the black hole physics,
it would also be important to investigate the evolution of the cosmological perturbations.
The linear cosmological perturbation theory and comparison of it with the observational data
of the type Ia suparnovae, baryon acoustic oscillations,
and cosmic microwave background have been argued in \cite{sarida}.
Then, the investigation of the linear and nonlinear growth of the matter density perturbations
may be able to differentiate scalar-tensor theories,
even though they could realize the very similar expansion history of the universe,
as has been investigated for the various scalar-tensor/modified gravity theories 
(see e.g., Refs. \cite{sf1,sf2,sf3,sf4,sf5,sf6}).
We hope to come back to these issues in our future publications.

\vspace{0.3cm}

{\it Note Added:}
While this paper was being completed,
Refs. \cite{sim1,sim2} on the similar topics have appeared on arXiv.
A part of the results presented in this paper have overlap with those in these papers.

\section*{ACKNOWLEDGMENT}
This work was supported by the FCT-Portugal through Grant No. SFRH/BPD/88299/2012.
We wish to thank the anonymous reviewer for his/her suggestions.
We also thank A. Flachi, C. Germani, L. Heisenberg, J. S. Lemos and E. N. Saridakis for comments.

%\newpage
\appendix

%%%%%%%%%%%%%%%%%%%%%%%%%%%%%%%%

\end{document}